\documentclass{mn2e}
\usepackage{graphicx}

\def \kms {${\rm{km}\,\rm{s}^{-1}}$}

\def \ixsig {index$-\sigma$}
\def \rcl {$R_{\rm cl}$}
\def \z {\phantom{0}}
\def \oiii  {[O\,{\sc iii}]}
\def \oiiib  {[O\,{\sc iii} $\lambda5007$]}

\def \hb   {H\,$\beta$}
\def \afe  {[$\alpha$/Fe]}
\def \zh  {[Z/H]}

\def \apjs{ApL}

\def \apj{ApJ}
\def \aj{AJ}
\def \mnras{MNRAS}
\def \ana{A\&A}

\def \z{\phantom{0}}
\title[The NOAO Fundamental Plane Survey -- III.]{The NOAO Fundamental Plane Survey --\\III. 
Variations in the stellar populations of red-sequence galaxies from the cluster core to the virial radius.}
\author[Russell J. Smith et al. ]
{Russell J. Smith$^{1,2}$\thanks{E-mail: russell.smith@durham.ac.uk},
Michael J. Hudson$^1$,
John R. Lucey$^2$,\newauthor
Jenica E. Nelan$^{3,4}$ and
Gary A. Wegner$^3$
\\
$^1$Department of Physics, University of Waterloo, 200 University Avenue West, Waterloo, Ontario \ N2L 3G1, Canada\\
$^2$Department of Physics, University of Durham, South Road, Durham, \ DH1 3LE, UK\\
$^3$Department of Physics and Astronomy, Dartmouth College, Hanover \ NH, USA\\
$^4$Department of Astronomy, Yale University, 260 Whitney Avenue, New Haven \ CT, USA
}
\date{Submitted to MNRAS, 8th February 2006}
\pagerange{\pageref{firstpage}$-$\pageref{lastpage}} \pubyear{2006}
\begin{document}
\label{firstpage}
\maketitle
\begin{abstract}
We analyse absorption line-strength indices for $\sim$3000 red-sequence 
galaxies in 94 nearby clusters, to investigate systematic
variations of their stellar content with location in the host cluster. 
The data are drawn from the NOAO Fundamental Plane Survey. Our adopted method is a generalization 
of that introduced by Nelan et al. to determine the global age--mass and 
metallicity--mass relations from the same survey. 
We find strong evidence for a change in galaxy properties, at fixed mass, over a range from the cluster 
centre to the virial radius, $R_{200}$. For example, red-sequence galaxies further out in 
the clusters have weaker Mgb5177 (at $\sim$8$\sigma$ significance) and stronger H$\beta$ and H$\gamma$
absorption ($\sim$3$\sigma$, $\sim$4$\sigma$) than galaxies of the same velocity dispersion in the 
cluster cores. The Fe5270 and Fe5335 indices show only very weak trends with radius. 
Using a total of twelve indices, the pattern of cluster-centric gradients is considered in
light of their different dependences on stellar age and chemical composition. 
The measured gradients for all twelve indices can be reproduced by a model in which red-sequence galaxies at 
$\sim$1\,$R_{200}$ have on average younger ages (by $15\pm4$\%) and lower $\alpha$-element abundance ratios 
(by $10\pm2$\%), than galaxies of the same velocity dispersion but located near the cluster centres.
For the total metallicity, Z/H, no significant gradient is found ($2\pm3$\% larger at $R_{200}$ than in the cores). 
There are hints that the age trend may be stronger for galaxies of lower mass and/or for galaxies with more
disky morphology. We show, however, that the trends cannot be driven primarily by changes in the
morphological mix as a function of radius. 
The cluster-centric age and \afe\ gradients are in the sense expected if galaxies in the cluster core
were accreted at an earlier epoch than those at larger radii, and if this earlier accretion contributed to 
an earlier cessation of star formation. The size of the observed age trend is comparable
to predictions from semi-analytic models of hierarchical galaxy formation. 
\end{abstract}
\section{Introduction}

Evidence that galaxy properties depend on their large-scale environment dates 
from at least seven decades ago (Hubble \& Humason 1931), with the morphology--density relationship
being the classic manifestation (e.g. Melnick \& Sargent 1977; Dressler 1980). 
The spectral properties of the galaxy population similarly exhibit 
environmental correlations, with more emission-dominated objects in the field than
in clusters. Measurements of H$\alpha$ equivalent widths in large spectroscopic 
surveys reveal a suppression of the average star-formation rate for galaxies in clusters, 
with this effect extending well beyond the virialized cluster core %$\sim$3 times the virial radius, $R_{200}$
(Lewis et al. 2002; G\'omez et al. 2003). 
A distinct but related issue is whether the correlation of star-formation rate with density is 
simply a consequence of the morphology--density relation, or whether there is a suppression
for cluster galaxies relative to field objects of the same morphological type (e.g. Balogh et al. 1998). 

A striking feature of the galaxy population within rich clusters is the sharp envelope
in colour--magnitude space defined by the reddest galaxies, and interpreted
as the locus of non-star-forming galaxies. Objects on this `red sequence' are 
predominantly of early-type morphology, i.e. ellipticals and S0s; 
their structural and dynamical parameters follow tight scaling relations such 
as the Fundamental Plane (e.g. Djorgovski \& Davis 1987), while their colours 
and spectroscopic line strengths reveal that their average stellar populations 
are correlated with the luminosity or mass (e.g. Visvanathan \& Sandage 1977). 
After star formation ceases, a galaxy evolves onto the red sequence within
a fairly short time interval of $\sim$1\,Gyr. This rapid reddening helps to 
explain the strong bimodality of colours in the overall 
galaxy population (e.g. Hogg et al. 2002). 

Beyond one Gyr after the end of star formation, the broadband optical colours become 
less sensitive to the stellar age; moreover, the effects of increased 
metallicity closely mimic the effects of greater age. This classic `age--metallicity 
degeneracy' allows the red sequence to conceal a wide range of ages, especially 
if there is a `conspiracy' such that galaxies with younger stars also have higher 
metallicities, at given mass. 
To resolve age differences within the red sequence, indicators which are more 
sensitive and more robust than broadband colours are needed. A widely-applied technique exploits the age 
dependence of the Balmer absorption lines, in combination with metallicity-driven indices, to break 
the age--metallicity degeneracy (e.g. Worthey 1994). A number of recent works
with this method have concluded that age must vary along the red sequence, with
less massive galaxies having younger ages than more massive systems,
(e.g. Caldwell et al. 2003; Nelan et al. 2005; Bernardi et al. 2006). 
These studies also support an increase of overall metallicity with mass, as well
as an increase in the abundance of $\alpha$ elements relative to iron. 

Beyond the existence of the overall age--mass trend, 
the distribution of red-sequence galaxy ages should reflect how star formation ended in galaxies 
in different environments. 
In N-body simulations of hierarchical clustering, the average accretion time of sub-haloes in 
cluster-sized parent haloes appears to be correlated with their final radius from the cluster centre (Gao et al. 2004, 
but see also Moore, Diemand \& Stadel 2004). In the simulations analysed by Gao et al., galaxies
within 20\% of the virial radius at $z=0$ entered the cluster at an average of $z\sim1$ (lookback time $\sim$8\,Gyr), 
while those at the virial radius were accreted only at $z\sim0.4$ ($\sim$4\,Gyr ago).
Any mechanism which suppresses star formation when a galaxy enters a cluster, combined with the accretion-time gradient, 
will yield a cluster-centric gradient in the stellar populations. As discussed by Balogh, Navarro \& Morris (2000), an attractive
candidate for this mechanism is `suffocation': the removal of the hot gas reservoir from the in-falling galaxy, 
causing a gradual decline in star-formation rate on time-scales of 1--3\,Gyr (see also Larson, Tinsley \& Caldwell 1980). 
If such a process acted on at least some of the galaxies now on the red sequence, a consequence will be for objects
now at larger radius to have experienced more extended star-formation histories on average. 
At any given mass, the outer galaxies will have luminosity-weighted ages slightly younger, and $\alpha$-element abundances  
(driven by the ratio of prompt SNe II versus delayed SNe Ia) slightly lower than galaxies near the core of the cluster. 
This behaviour is now captured qualitatively by semi-analytic models of galaxy formation, which combine numerical prescriptions for star formation, 
and other baryonic physics, with N-body dark-matter merger trees. For instance, in the models of De Lucia et al. (2006), the mean
luminosity-weighted age of bulge-dominated galaxies falls from $\sim$12\,Gyr at cluster centres to $\sim10$\,Gyr at the virial 
radius. Although some of this effect is due to mass segregation (since the largest galaxies are slightly older, and
preferentially located near the centre),  this is not the dominant cause, and a true residual trend of age with radius, 
at fixed mass, appears to be present. 

While there is substantial observational evidence for truncation of star formation in clusters for the galaxy population
at large, there is less conclusive evidence for environment-driven effects in the red galaxies themselves. 
In this case, since the mass-sequence of galaxies dominates most of the observed correlations, any environmental effect would be revealed by 
systematic shifts in the residuals from the overall scaling relations, in particular the colour--magnitude relations 
(e.g. Terlevich, Caldwell \& Bower 2001; Hogg et al. 2004)  or 
the spectroscopic line index versus velocity dispersion (index$-\sigma$) relations. 
Comparing samples drawn from field and cluster environments, Bernardi et al. (1998) found marginally smaller Mgb5177, at 
given velocity dispersion, for their field sample. Kuntschner et al. (2002) used a range of indices to infer that very isolated
early-type galaxies have younger stellar ages than their counterparts in clusters. 
An alternative to the cluster-versus-field approach is to compare the properties of cluster galaxies at different radii from 
the cluster centre. An early study of this kind was made by Guzm\'an et al. (1992), 
who found a significant offset in the magnesium index Mg$_2$, between the core of Coma and galaxies at $\sim1.5\,R_{200}$. 
The results were interpreted as evidence for a gradient in stellar ages, with more recent star-formation in the halo sample. 
Also in Coma, Carter et al. (2002) have claimed to find an offset in the same sense for Mgb5177, but only very weak gradients
in the age-sensitive H$\beta$ and the iron-abundance indicator $\langle{}$Fe$\rangle=\frac{1}{2}({\rm Fe5270}+{\rm Fe5335})$. 
Considering the relative strength of the effect
in the three indices, Carter et al. favour an explanation in terms of a cluster-centric metallicity gradient, 
and invoke pressure confinement of supernova ejecta by intra-cluster gas as the underlying mechanism. 

In this paper we determine the cluster-centric gradients in the index$-\sigma$ residuals, using data from the National Optical Astronomy Observatory (NOAO)
Fundamental Plane Survey (NFPS). We use the gradients for a set of twelve spectral indices, with different responses to the stellar population parameters, 
to make a clear separation of age and metallicity effects. The method applied is a generalisation of that used by Nelan et al. (2005)
to investigate the trends of stellar populations with increasing mass, based on the same sample. The layout of the present paper is as follows: 
In Section~\ref{sec:data} we first describe the sample and data employed, while in Section~\ref{sec:oneix} we determine
the radial trends followed by twelve individual line-strength indices. These results are independent of any adopted stellar population models. 
In Section~\ref{sec:interpret}, we use models of Thomas et al. (2003, 2004) with the aim of describing the 
radial behaviour of all twelve of the indices, in terms of underlying trends in age, metallicity and 
$\alpha$-element enhancement. We consider in some detail the degree to which a morphology--radius trend contributes
to these results, and present extensive tests for systematic biases.
Section~\ref{sec:discuss} compares the results obtained against previous work within clusters, and with cluster-versus-field 
studies. We also consider the physical origin of cluster-centric age gradients in hierarchical structure formation models, 
and compare our results with predictions from semi-analytic modelling. Finally, our principal conclusions are
reiterated in Section~\ref{sec:concs}.

\section{Data}\label{sec:data}

The data to be used in this analysis are fully described in Smith et al. (2004), which also presents an overview of the NFPS programme,
and in Nelan et al. (2005), where the line-strength measurements are reported in detail. We refer the reader to these papers for a full discussion; 
here we provide only a summary. 

The measurements are based on multi-fibre spectroscopic data obtained at the  Cerro-Tololo Inter-American Observatory (CTIO) 4-m telescope
and the Wisconsin--Indiana--Yale--NOAO (WIYN) 3.5-m telescope. Absorption line-strengths, as defined by Trager et al. (1998) and Worthey \& Ottoviani (1997),
were measured after smoothing the spectra to match the resolution of the Lick stellar library, in order to match predictions 
based on the Lick fitting functions. Unlike some studies, we made no attempt to bring the data onto the Lick system by comparing 
standard stars. Rather, the indices are measured on flux-calibrated spectra. As will be seen, our method employs only relative changes 
in the line-strengths, so absolute zero-point shifts from the Lick system would not affect our results. We do not in this paper apply aperture 
corrections to the line-strengths based on galaxy distance, preferring to absorb such effects into a free parameter in our fits. For the 
velocity dispersions, we apply an aperture correction of 0.04\,dex per decade in angular diameter distance 
(e.g. J\o rgensen, Franx \& Kj\ae rgaard 1995). 

The set of twelve indices analysed here is the same as used by Nelan et al. to determine the mass scaling relations. As in the previous paper, 
we will interpret the index data in the context of models by D. Thomas and collaborators (Thomas, Maraston \& Bender 2003; 
Thomas, Maraston \& Korn 2004; jointly hereafter TMBK). These models are widely used since they incorporate the effects
of abundance ratio variations, through the parameter \afe. The super-solar \afe\ measured in giant elliptical galaxies (e.g. Worthey, Faber \& Gonz\'alez 1992) 
are interpreted as evidence for star-formation time-scales which are short compared to the $\sim$1\,Gyr lifetimes of Fe-producing SNIa. 
More recently, abundance ratio effects have been incorporated into alternative models (e.g. Lee \& Worthey 2005; Schiavon 2005); 
for ease of comparison, however, we use only the TMBK models in this paper.  
There is an element of subjectivity in the choice of indices to incorporate in the fits, but the index set needs to be 
sensitive to the three parameters covered by the models, i.e. age, metallicity and $\alpha$-abundance. Our chosen set includes three 
$\alpha$-element indicators (Mgb5177, CN1, Fe4668\footnote{Sometimes called C$_2$4668 because of its strong dependence on the C 
abundance, see Tripicco \& Bell (1995).}), and three age-sensitive Balmer indices (HdF, HgF, Hbeta)\footnote{
We will adopt the convention that HdF etc refer to the line-strength indices, defined by fairly broad continuum and central 
pass-bands, while H$\delta$ etc refer to physically meaningful spectral lines, either in emission or absorption.}. The remaining six indices 
measure primarily the iron-peak metallicity (Fe4531, Fe5270, Fe5335, Fe5406, Fe5709, Fe5782). The balance between the three groups is 
intended to compensate for the intrinsic sensitivity of the indices: the Fe indices only depend weakly on the underlying parameters, 
relative to the typical errors, so a greater number should be included to reach a given significance level. 

Because the NFPS galaxy sample covers a broad redshift range, $0.01\la{}z\la0.08$, the strong night-sky lines at 5577\,\AA\ and 5895\,\AA\
preclude accurate line-strength measurements for indices redder than Mgb5177, over certain redshift intervals. For instance Fe5270
cannot be measured for $cz$ between 14500\,\kms\ and 20000\,\kms. A feature of our analysis method, described below, is that we do 
not require an identical set of indices to be measured for all galaxies in the sample. Using multiple Fe line-strengths in particular allows
the whole redshift range to be covered by multiple semi-redundant indices. In the case of the reddest indices, Fe5709 and Fe5782, 
we further restrict the redshift range such that measurements at observed wavelengths longer than 6050\,\AA\ do not enter the analysis. This region
corresponds to the extreme red part of the spectrum in the CTIO data, and may be subject to flux-calibration problems. 
Note that for consistency between northern and southern data subsets, we reject all line-strengths beyond this cut-off, including those 
measured from WIYN spectra.

Although the line-strength indices are designed to measure stellar absorption features in integrated spectra, some of the indices are 
contaminated by the presence of nebular emission lines at the same wavelengths. The most troublesome cases are for the age-sensitive 
Balmer indices, contaminated by nebular emission at H$\beta$, H$\gamma$ and H$\delta$. The effect of emission is to `fill in' the 
stellar absorption, leading to erroneous age estimates which are older than would be determined in the absence of emission. Relative 
to the age-effect in the stellar spectrum, the nebular contamination is more severe for H$\alpha$ and H$\beta$ than 
for the higher-order lines. With data of very high quality this can in principle be made the basis of an emission-correction scheme 
(e.g. Caldwell, Rose \& Concannon 2003). For NFPS, however, only a small subset of galaxies have H$\alpha$ measurements 
(see Smith 2005). 
More traditionally, emission corrections have assumed a constant ratio of \oiii\ to \hb\ equivalent widths, and used the easily-measured \oiii\
line at 5007\AA\ to estimate the contamination at \hb\ (e.g. Gonz\'alez 1993). In Nelan et al. (2005) we 
showed, however, that the \oiii\ and \hb\ data for the NFPS sample show a wide (factor of $\sim$20) variation in line ratios. Moreover 
the distribution of line ratios is bimodal (see also Yan et al. 2005), and strongly related to the galaxy mass or luminosity (as also discussed by Goudfrooij 1999). 
As a result, the frequently-adopted factor of $0.6\times$\oiiib\ (Trager et al. 2000) is a poor and misleading correction, even in a statistical sense.
In the present paper we apply a more conservative approach, selecting the galaxy sample to exclude objects with nebular emission according 
to the measurements of Nelan et al. Specifically, we adopt selection limits of $EW(\rm H\beta)>-0.6$\AA\ and $EW$\oiiib$>-0.8$\AA\ 
(where the equivalent widths $EW$ are negative for emission, as tabulated by Nelan et al.), 
and galaxies not meeting {\it both} of these criteria are removed from the sample entirely. 
Thus all of our results will strictly refer to emission-free red-sequence galaxies. 

For an explicitly cluster-based survey such as NFPS, the projected cluster-centric distance, \rcl, is a natural and convenient (although imperfect)
parameter to describe the environment of a galaxy. We adopt the nominal cluster centres as reported by Smith et al. (2004), which were based on
either the position of one or more dominant galaxies, or a clear X-ray emission peak, on a case-by-case basis. To account for the
range in richness of the NFPS clusters, we express \rcl\ in terms of the cluster virial radius, $R_{200}$, defined as
\begin{equation}
R_{200} = \sqrt{3} \left( \frac{\sigma_{\rm cl}}{1000\,{\rm km\,s}^{-1}} \right)\,h^{-1}\,{\rm Mpc}
\end{equation}
where $\sigma_{\rm cl}$ is the cluster velocity dispersion (e.g. Carlberg et al. 1997) and $h$ is the Hubble constant in units of 100\,\kms\,Mpc$^{-1}$.
We note that alternative definitions have been used for the virial radius by other authors (e.g. Quintero et al. 2005), but in the present paper
we are concerned primarily with placing any gradients on a relative scale, and to first order the above should be sufficient. 

Finally, since this work will investigate radial trends in the properties of `normal' red-sequence galaxies, we need to be
careful in our treatment of cD galaxies. Since cDs are found preferentially near the cluster centre but probably have distinctive star-formation
histories relative to other giant elliptical galaxies (e.g. Crawford et al. 1999), they could exert an undue influence on our fits. 
However, some cDs in our sample are not coincident with the nominal cluster centres as listed by Smith et al. (2004), so 
an inner radial cut is not sufficient to remove all of them from the sample. Instead, we use the J-band luminosity from 
the Extended Source Catalogue of the Two Micron All Sky Survey (Jarret et al. 2000), to exclude from the fit $\sim$200 galaxies with 
$M_J < -23.25$ (for $h=1$). For the isophotal magnitudes used for the selection, this limit is approximately one magnitude brighter than $M^\star_J$ (Kochanek et al. 2001; Cole et al. 2001). Although the cut will remove some normal giant ellipticals from the sample, as well as the cDs,
this should not bias the fits; the sample size is reduced by only $<$10\%, causing minimal loss in the overall signal-to-noise ratio.

\section{Modelling the index$-\sigma$ relations}\label{sec:oneix}

In this section, we fit linear multivariate models for the index measurements, including a term describing cluster-centric
variations at given mass. The results of this section are based only on observed variations in the line-strengths and hence do 
not depend on any particular stellar population model used to interpret the index data. 

\subsection{Motivation}

\begin{figure*}
\includegraphics[angle=270,width=180mm]{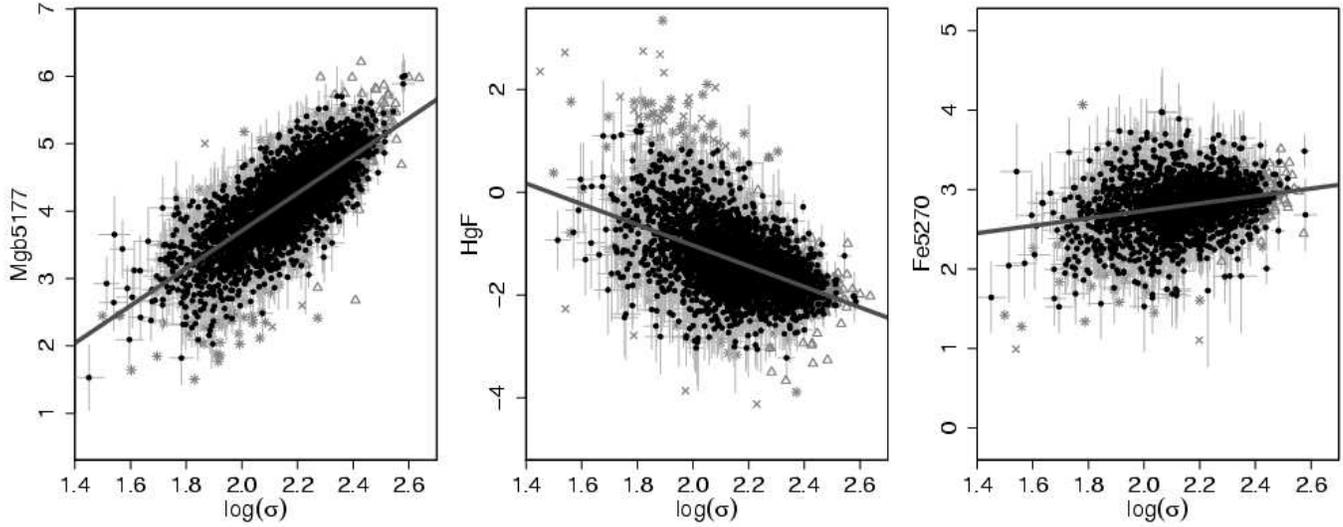}
\caption{The index--$\sigma$ data for three illustrative indices, Mgb5177, HgF and Fe5270. The index measurements have been adjusted for \rcl\ and
$cz$ following the best fitting models from Section~\ref{sec:fitresults}, although in practice this is a small correction. The galaxies used in the final fit are denoted
with black points and grey error bars. Grey symbols without error bars indicate objects not used in the fit: asterisks are galaxies rejected
due to nebular emission lines; triangles are objects rejected because they have cD-like J-band luminosities; crosses indicate galaxies
removed by the iterative outlier rejection scheme. Note that the fitted objects are plotted last, thus only the outlying grey points are visible.
The best fitting relation is shown by the thick grey line.} 
\label{fig:rawixsig}
\end{figure*}

Before presenting the fits in detail, we first illustrate the presence
of cluster-centric gradients using three example indices, Mgb5177, HgF and
Fe5270. Figure~\ref{fig:rawixsig} shows the index--$\sigma$ data
for these indices, with the global relations as derived in the following 
section. As shown in Nelan et al., the relative slopes of the index--$\sigma$
relations can be understood in terms of systematic variations in stellar age and chemical
composition along the mass sequence.

\begin{figure*}
\includegraphics[width=107mm, angle=270]{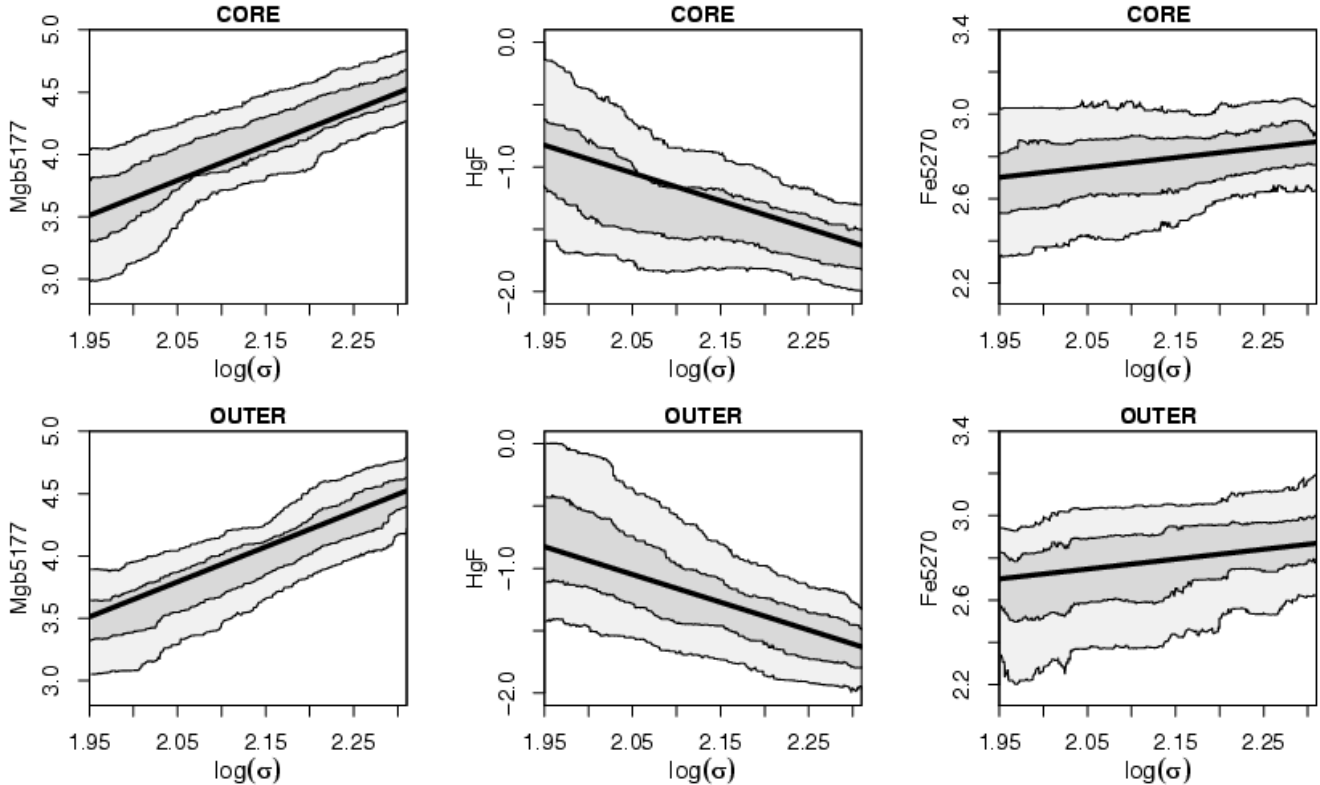}
\caption{Running quantiles of the Mgb5177$-\sigma$, HgF$-\sigma$ and Fe5270$-\sigma$ relations, for two extreme ranges of cluster-centric radius. The
`core' region (upper panels) is defined by $R_{\rm cl}/R_{200}<0.2$, while the `outer' region (lower panels) has $0.5<R_{\rm cl}/R_{200}<1.0$.
In each panel the pale and dark shaded regions enclose respectively two-thirds and one-third of the data in a sliding sample of 300 galaxies 
(hence the reduced range in $\sigma$ relative to Figure~\ref{fig:rawixsig}). 
The global linear \ixsig\ fit is reproduced
as the bold line for comparison. There is a clear offset of the Mgb5177$-\sigma$ relation between the core and outer samples. For 
HgF, the shift appears more pronounced at low masses, while in Fe5270 little difference is observed.}
\label{fig:quantrun}
\end{figure*}

As initial evidence for cluster-centric variations, Figure~\ref{fig:quantrun} shows quantiles of the \ixsig\ relations 
in two ranges of \rcl, corresponding to the cluster core ($R_{\rm cl}/R_{200}<0.2$), and to the outermost galaxies observed
($0.5<R_{\rm cl}/R_{200}<1.0$). The global relations are reproduced in each panel for comparison. 
The shift in the Mgb5177$-\sigma$ relation is readily seen, and is well described by a simple offset at all $\sigma$, 
with the outer sample showing smaller Mgb5177 values at given $\sigma$. 
Similarly, the HgF$-\sigma$ relation clearly differs between the two radial samples, with the outer sample having larger
HgF on average. In this case there is a suggestion of curvature in the relation for the core region, with the radial effect more pronounced 
in the low-mass galaxies ($\log\sigma\la2.1$). Finally, the Fe5270$-\sigma$ relation shows no discernible change with radius. 

Although in principle Figure~\ref{fig:quantrun} could form the basis of an radial-trends analysis,
this method would be quite sensitive to the boundaries adopted for the radial bins. Instead, 
we prefer a more flexible approach, treating \rcl\ as a continuous variable, and including
an extra term in the index--$\sigma$ relations, to describe the 
radial behaviour of the line-strengths. 

\subsection{Method}\label{sec:fitmethod}

We aim to describe the measured line-strength indices $I$, through a linear model of the form
\begin{equation}\label{eqn:ixsigmod}
I = a_0 + a_\sigma \log\sigma + a_z \log \frac{D_{\rm ang}(\rm cz_{CMB})}{D_{\rm ang}(cz_0)} + a_R\frac{R_{\rm cl}}{R_{200}}\, .
\end{equation}
The first two coefficients, $a_0$ and $a_\sigma$, represent the index--$\sigma$ slope and zero-point as usual. We assume throughout this paper
that the velocity dispersion, $\sigma$, can be regarded as a proxy for galaxy mass, and moreover that the relationship between 
$\sigma$ and mass is universal. 
The additional model coefficients, $a_z$ and $a_R$, respectively test for correlations with redshift, and with radius from cluster centre.
The principal objective of this paper is to constrain and interpret the $a_R$, which encode any systematic variation of the 
line-strengths with location in the cluster, after controlling for the overall trend with mass. 
Thus, a non-zero value of $a_R$ would {\it not} result from the largest galaxies being in the central 
part of the cluster. Rather, $a_R\neq0$ indicates a systematic environmental dependence in the properties of galaxies at a {\it given} mass.
In Section~\ref{sec:subsample} we discuss the effect of using a term proportional to $\log R_{\rm cl}/R_{200}$, instead of the linear
form above. 

\begin{figure}
\includegraphics[angle=270,width=85mm]{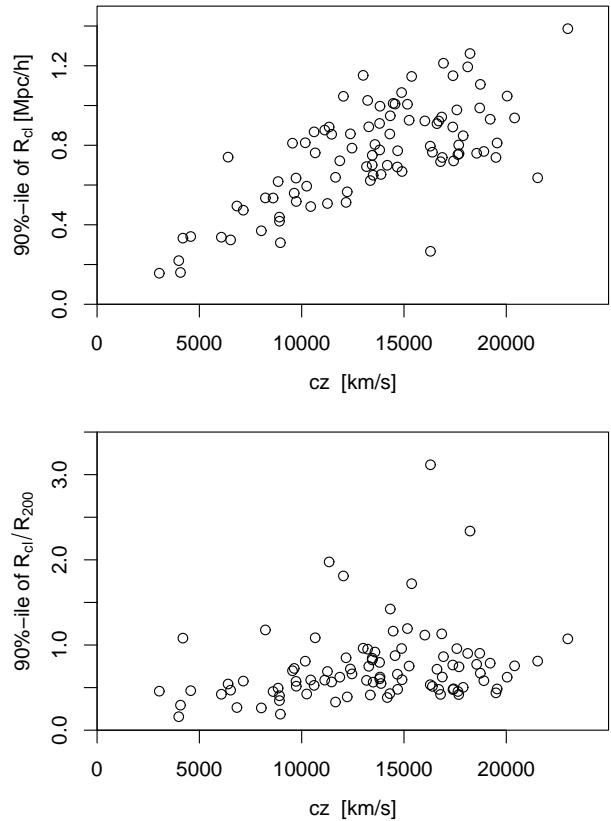}
\caption{Radial extent of the galaxy sample in each cluster, as a function of redshift. The upper panel shows that the more distant clusters in NFPS 
are sampled to a larger physical radius. In the lower panel we show the sample extent after scaling to the virial radius of each cluster. }
\label{fig:sample}
\end{figure}

Figure~\ref{fig:sample} shows the sample extent in each cluster, characterized by the radius $R_{90\%}$ which encloses 90\% of the sample galaxies, 
in terms of physical units (upper panel) and after scaling by the virial radius (lower panel). 
Given that the samples were collected with spectrographs of a fixed field of view (diameters 60\,arcmin at WIYN, 
38\,arcmin at CTIO), the physical area surveyed in a cluster is generally larger for more distant clusters than for those 
nearby. However, when scaled in terms of the virial radius, the sample extent is less tightly correlated with redshift. This arises 
partly because the cluster sample includes generally more massive systems (larger $R_{200}$) at greater distances, and partly because 
there is substantial range in the virial radius $R_{200}$ at any given distance. The cluster with $R_{90\%}/R_{200}>3$ is A3528B, which 
formally has velocity dispersion 50\,\kms, based on only six galaxies. Although such cases clearly yield unreliable $R_{200}$ values, 
by definition they contribute very few galaxies to the fits. Moreover, we restrict the model to radii $R_{\rm cl}/R_{200}<1$, 
so that such points do not carry undue influence (this cut excludes $\sim$4\% of the original sample). 

The parametrization of the redshift term is chosen to match the traditional form of the `aperture correction' 
(e.g. J\o rgensen et al. 1995). Here we adopt 10000\,\kms\ for the normalizing redshift $cz_0$; the redshift $cz$ is the CMB-frame velocity 
of the cluster to which the galaxy is assigned, as tabulated by Smith et al. (2004). We use the ratio of angular-diameter distance $D_{\rm ang}(cz)$, 
for a $\Omega_{\rm m}=0.3, \Omega_\Lambda=0.7$ cosmology, rather than the redshift ratio, but this makes little difference in practice. 
The $a_z$ coefficient allows for stellar-population gradients internal to the galaxies (e.g. Mehlert et al. 2003), which translate into 
redshift correlations as the observed aperture in kpc scales with distance. Including the redshift term allows the 
results properly to incorporate uncertainties in the aperture corrections, and any other systematic effects
which vary slowly with redshift. In particular, any low-order errors in flux calibration will cause weak
redshift dependence in the indices with `wide' definitions, e.g. CN1, Fe4668). For the most distant clusters, we not only observe a 
larger physical radius in each galaxy, but also we typically sample the cluster population out to larger radii. This introduces a covariance between 
$a_R$ and $a_z$ which cannot be ignored if meaningful confidence limits are to be determined.

To summarize the fitting method, our analysis uses only the emission-free galaxies within one virial radius, excluding those with cD-like J-band
luminosities. An iterative outlier-rejection scheme is employed, clipping galaxies deviating from the current model by four times
the total measured scatter. The number of galaxies in the fit depends on the particular index, since no measurements were made 
for indices contaminated by night-sky lines. In practice, $\sim$3000 galaxies are fit for the blue indices up to and 
including Mgb5177, while $\la$2000 galaxies are fit for Fe5270 and redder indices. 

Our modelling strategy minimises the residuals in the line-strengths, implicitly assuming that the index errors dominate over the 
uncertainty in $\log\sigma$. Such a fit will be unaffected by explicit selection in $\sigma$, or in the other predictor variables
$cz$ and \rcl. In practice, however the galaxy sample was initially selected on R-band total apparent magnitude, and the spectroscopic
success rate is higher for galaxies of higher central surface brightness. This could introduce a slight
bias towards observing younger (hence brighter) galaxies at any given mass, but one which should not be manifest at any
particular radius, and so should not affect the principal results of this paper.

\subsection{Results}\label{sec:fitresults}

\begin{figure*}
\includegraphics[angle=0,width=175mm]{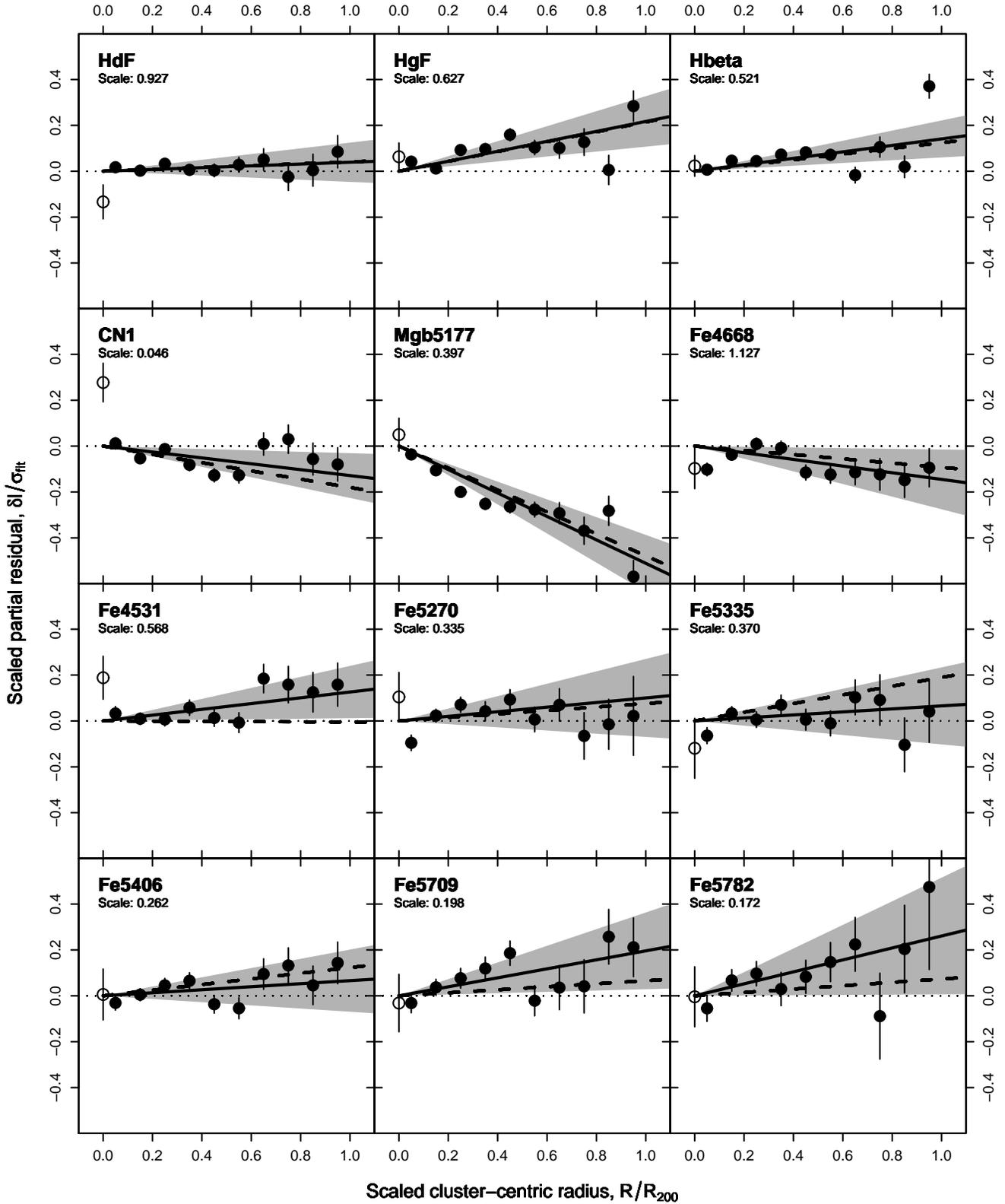}
\caption{Index-residual versus radius relations. Solid points with error bars show the weighted mean of the partial residuals $\delta{}I$, defined as in 
Equation~\ref{eqn:partres}, binned in annuli of constant width 0.1\,$R_{200}$. The $\delta{}I$ represent the difference between the measured index and 
that predicted from the velocity dispersion and redshift only, as a function of cluster-centric radius. The residuals have been scaled by the total 
observed scatter in the \ixsig\ relation, the scale factor being noted in each panel. 
The open symbol at \rcl=0 refers to centrally-located cD candidates, which were not used in the fits. 
The heavy solid line shows our linear fits, with the 2$\sigma$ range of $a_R$ indicated by 
a shaded wedge. Finally, the heavy dashed line shows the slope predicted under our best-fitting model
of Section~\ref{sec:radparvar}, based on a linear radial variation in galaxy age, metallicity and
\afe.}
\label{fig:partreds}
\end{figure*}

\begin{table*}
\caption{Coefficients of the single-index fits. The total number of galaxies for which the index is measured
is given as $N_{\rm gal}$. After applying selection criteria discussed in the text, the sample for each index
is reduced to $N_{\rm sel}$. In the course of the fitting, a 4$\sigma$ residual clipping is applied, so that 
the final fit is for $N_{\rm fit}$ objects. The coefficients  $a_0,a_\sigma,a_z,a_R$ are as in 
Equation~\ref{eqn:ixsigmod}. For the coefficient of cluster-centric radius, $a_R$, the significant ($2\sigma$) cases
are highlighted with an asterisk. The final column gives the ratio of mass-trend to radius-trend, as discussed in 
the text.}
\label{tab:fitcoeffs}
\begin{tabular}{lcccccccc}
\hline
Index & $N_{\rm gal}$ & $N_{\rm sel}$ & $N_{\rm fit}$ & $a_0$ & $a_\sigma$ & $a_z$ & $a_R$ & ratio $a_R/a_\sigma$\\
\hline
HdF        & 3765 & 3177 & 3146 & $+3.362\pm0.135$ & $-1.352\pm0.060$ & $+0.107\pm0.061$ & $+0.036\pm0.040^{\phantom{\star}}$ & $-0.027\pm0.029$ \\
HgF        & 3765 & 3177 & 3152 & $+2.987\pm0.120$ & $-2.010\pm0.053$ & $+0.172\pm0.051$ & $+0.136\pm0.034^\star$ & $-0.068\pm0.017$ \\
Hbeta      & 3765 & 3177 & 3169 & $+4.397\pm0.073$ & $-1.218\pm0.032$ & $+0.260\pm0.030$ & $+0.073\pm0.021^\star$ & $-0.060\pm0.017$ \\
CN1        & 3765 & 3177 & 3145 & $-0.333\pm0.008$ & $+0.179\pm0.003$ & $-0.024\pm0.004$ & $-0.006\pm0.002^\star$ & $-0.033\pm0.013$ \\
Mgb5177    & 3716 & 3142 & 3139 & $-1.849\pm0.085$ & $+2.780\pm0.037$ & $-0.227\pm0.036$ & $-0.203\pm0.025^\star$ & $-0.073\pm0.009$ \\
Fe4668     & 3765 & 3177 & 3165 & $-4.090\pm0.248$ & $+4.735\pm0.110$ & $-1.381\pm0.112$ & $-0.163\pm0.073^\star$ & $-0.035\pm0.016$ \\
Fe4531     & 3765 & 3177 & 3160 & $+1.537\pm0.110$ & $+0.779\pm0.049$ & $-0.260\pm0.047$ & $+0.072\pm0.032^\star$ & $+0.092\pm0.042$ \\
Fe5270     & 2152 & 1824 & 1822 & $+1.797\pm0.087$ & $+0.468\pm0.038$ & $+0.077\pm0.042$ & $+0.034\pm0.028^{\phantom{\star}}$ & $+0.072\pm0.061$ \\
Fe5335     & 2291 & 1966 & 1962 & $+1.368\pm0.096$ & $+0.554\pm0.043$ & $-0.053\pm0.037$ & $+0.024\pm0.031^{\phantom{\star}}$ & $+0.044\pm0.056$ \\
Fe5406     & 3071 & 2583 & 2580 & $+0.990\pm0.063$ & $+0.309\pm0.028$ & $-0.102\pm0.023$ & $+0.017\pm0.018^{\phantom{\star}}$ & $+0.056\pm0.058$ \\
Fe5709     & 1831 & 1524 & 1522 & $+1.133\pm0.055$ & $-0.143\pm0.024$ & $+0.007\pm0.022$ & $+0.039\pm0.017^\star$ & $-0.270\pm0.124$ \\
Fe5782     & 1047 &  888 &  887 & $+0.211\pm0.065$ & $+0.239\pm0.029$ & $+0.091\pm0.047$ & $+0.045\pm0.022^\star$ & $+0.188\pm0.094$ \\
\hline
\end{tabular}
\end{table*}

The linear fit results are summarized in terms of the \ixsig\ relation coefficients, in 
Table~\ref{tab:fitcoeffs}. 

For the three indices Mgb5177, HgF and Fe5270, the global \ixsig\ relations were shown in 
Figure~\ref{fig:rawixsig}. Although at first sight the slope of the fitted relations appears by eye to be somewhat 
shallow compared to the data, a careful examination confirms that the fit approximately bisects the {\it vertical} distribution
at any given $\sigma$, and thus is a good predictor of the line-strengths, as required. 

Figure~\ref{fig:partreds} presents the results on radial 
trends for all twelve indices. Each panel shows the `partial residuals' of the corresponding 
\ixsig\ relation, i.e.
\begin{equation}\label{eqn:partres}
\delta I = I - \left[ a_0 + a_\sigma \log\sigma + a_z \log \frac{D_{\rm ang}(\rm cz_{CMB})}{D_{\rm ang}(cz_0)} \right]
\end{equation}
averaged in bins of \rcl. The aim is to isolate the variation associated with cluster-centric 
radius trends after correcting for the mass and redshift terms. Comparing to Equation~\ref{eqn:ixsigmod}, it is clear that 
$\delta I = a_R R_{\rm cl}$, if the linear model holds. The figure shows our 2$\sigma$ limits on $a_R$ as a shaded wedge, 
for comparison to the binned residuals. If the radial behaviour is more complex than 
a linear trend, this will be revealed in a poor match between the partial residuals and the fitted $a_R$. 

\begin{figure*}
\includegraphics[width=120mm, angle=270]{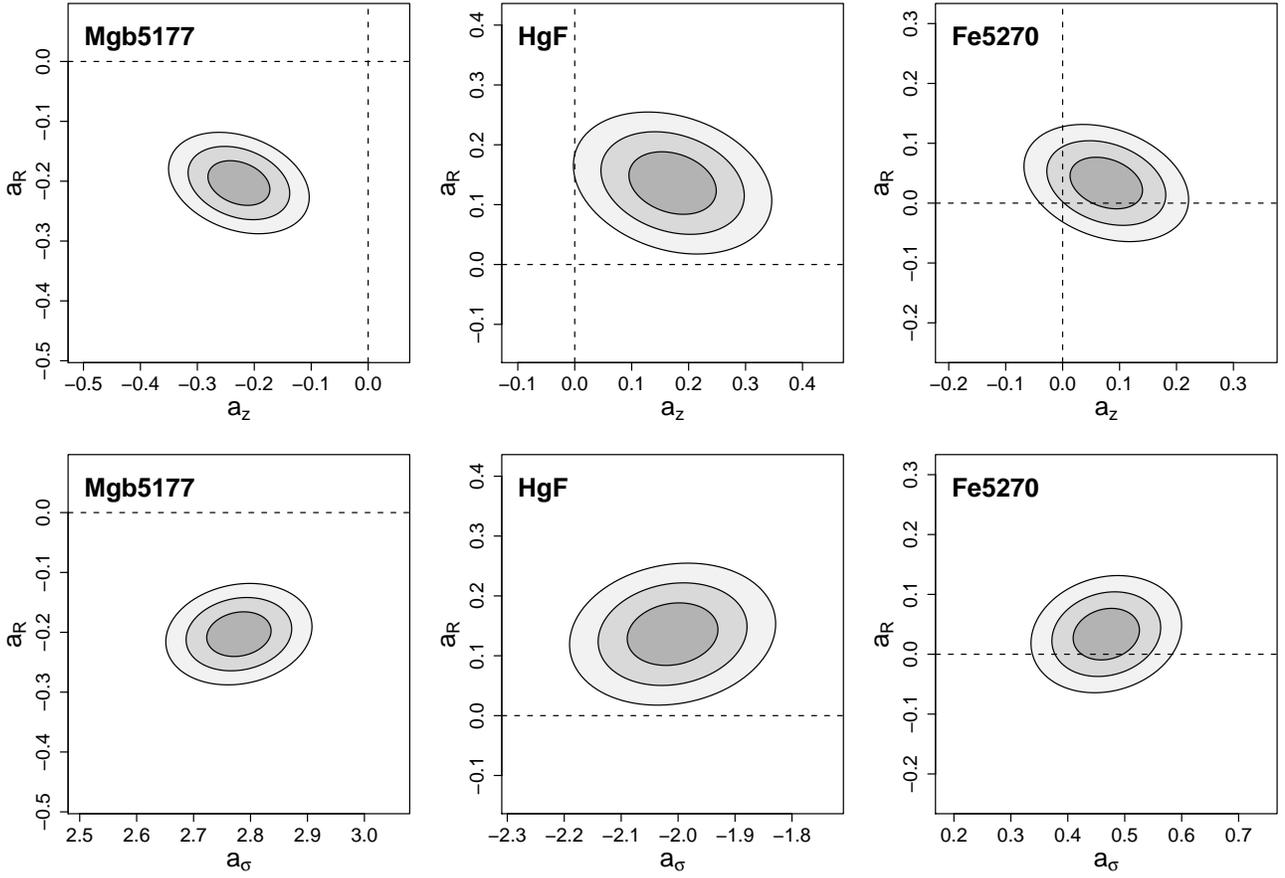}
\caption{Error ellipses for the single-index fits, for the example cases of Mgb5177, HgF and Fe5270. The upper row shows the 1,2,3-$\sigma$ confidence 
regions for the coefficients of cluster-centric radius and redshift in Equation~\ref{eqn:ixsigmod}. In the lower rows, the equivalent 
regions are shown for the coefficients of radius and $\log\sigma$. Dashed lines indicate zero for reference. 
}
\label{fig:egellipses}
\end{figure*}

For our three illustrative indices, we show the error ellipses in Figure~\ref{fig:egellipses}, 
demonstrating the modest covariance between $a_R$ and the other coefficients. The correlation coefficient 
for the $a_R$ and $a_z$ parameters is $-0.32<r<-0.24$ with larger values for the red 
Fe lines which are available over a narrower range of redshift due to sky-line contamination and the more
limited spectral extent at CTIO. The cause of this effect is the correlation of sample radial extent
with redshift, as discussed above. There is also a small covariance between $a_R$ and $a_\sigma$ ($0.13<r<0.19$), 
in the sense that fitting a flatter slope with respect to mass causes a steeper radial gradient. This cross-talk 
arises because sample galaxies near the cluster
centre have a larger $\sigma$ on average than those further out ($\sim$0.1\,dex out to the virial radius), which
could be due to true mass segregation in the cluster, but could also be an artefact of greater target density
in the centre, coupled with our prioritizing brighter targets for spectroscopic follow-up. In practice, the level of covariance
is small enough that reasonable changes ($\sim$10\%) in $a_\sigma$ do not significantly influence the radial trend.

Taken jointly, the overall significance of the \rcl\ trends is very high. With 12 indices we would expect of order
four of the $a_R$ coefficients to be non-zero at $1\sigma$ by chance, including perhaps one at $2\sigma$. In fact, the Mgb5177
trend is significant at $\sim$8$\sigma$, while  several others are at $>3\sigma$. 
Relative to a model in which the indices are uncorrelated with cluster-centric radius, we formally have
$\chi^2\sim126$  % sum((ModelData$slope.rad / ModelData$error.rad)^2)
on 12 degrees of freedom. We can therefore 
already exclude the null hypothesis and conclude with high degree of confidence that the 
line-strengths of red-sequence galaxies in the NFPS dataset exhibit signicant variations as a 
function of cluster location.

\begin{figure*}
\includegraphics[width=70mm, angle=270]{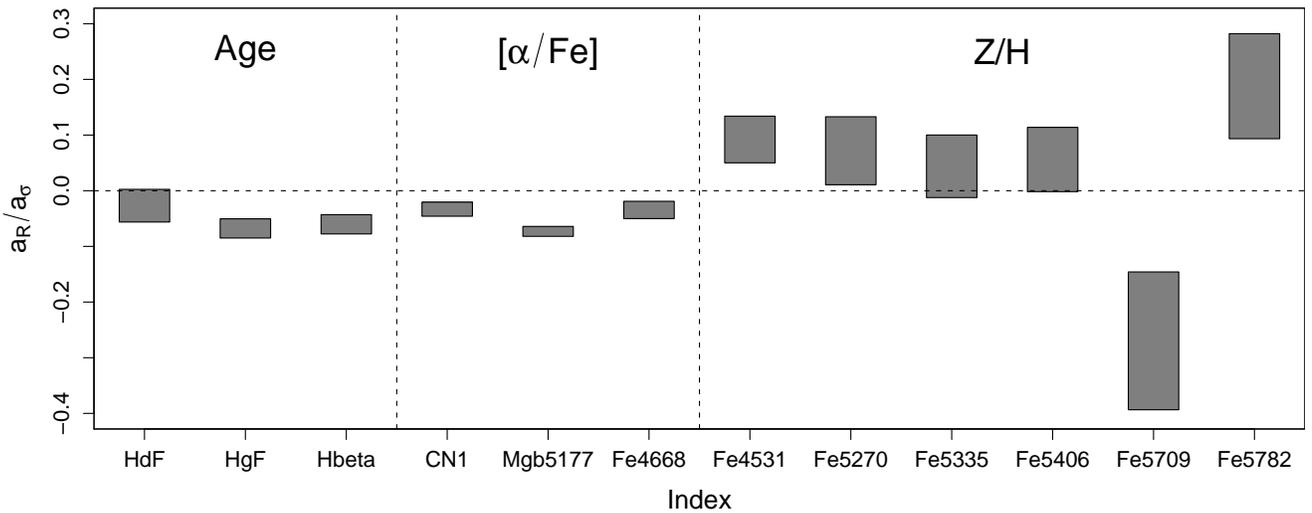}
\caption{Ratio of radial-slope to \ixsig\ slope, i.e. $a_R/a_\sigma$. The shaded boxes indicate the $1\sigma$ limits on this ratio
as given in Table~\ref{tab:fitcoeffs}. The vertical bars divide the indices into subsets 
broadly reflecting their responses to age, metallicity and $\alpha$/Fe ratio, in the models of TMBK. The significant index-to-index variation
in $a_R/a_\sigma$ argues for an environmental mechanism which is distinct from the overall mass trend (see text). 
}
\label{fig:radsigrat}
\end{figure*}

\begin{figure*}
\includegraphics[angle=270,width=180mm]{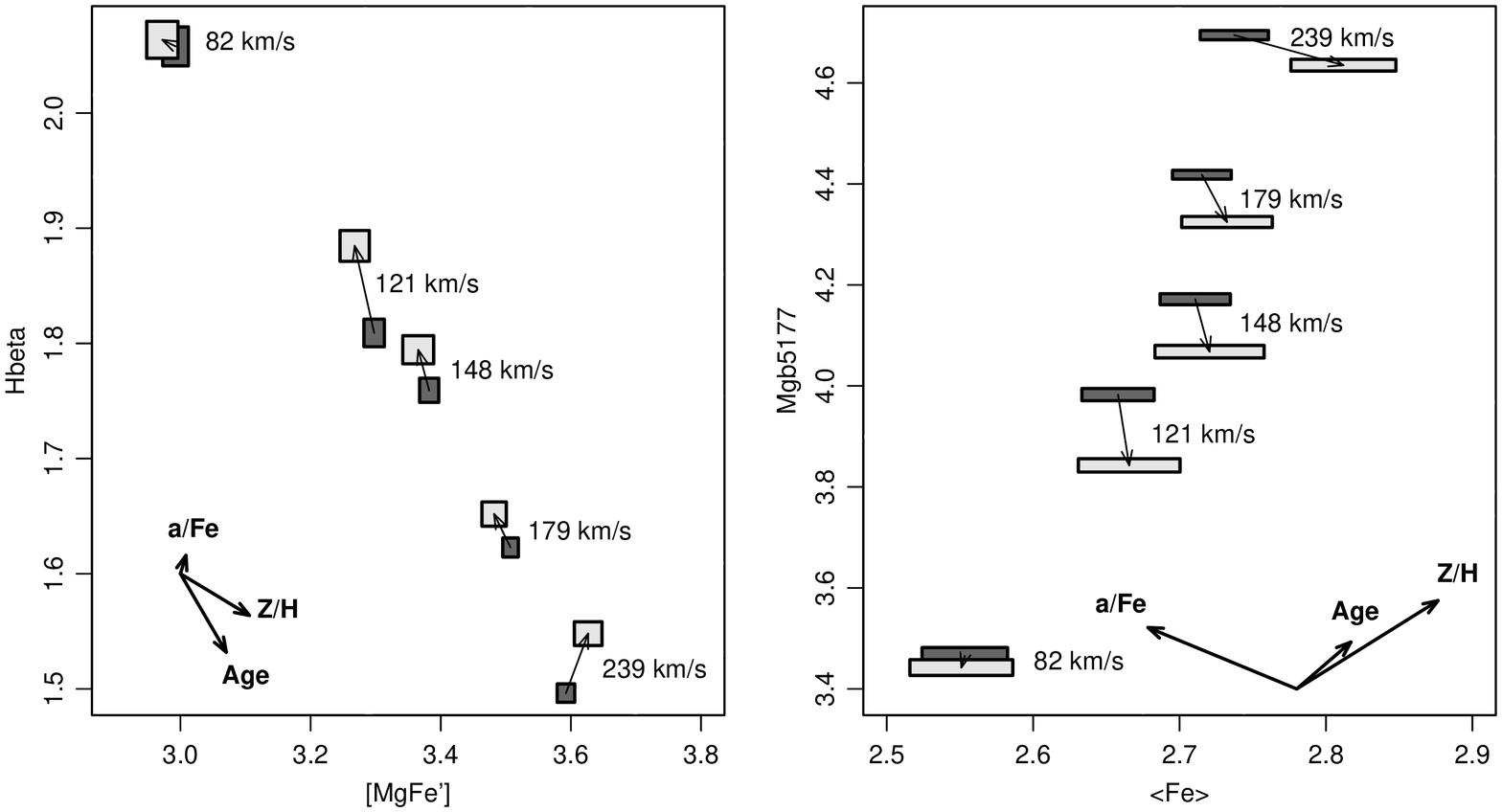}
\caption{
Classic two-index diagrams. The left panel shows Hbeta versus [MgFe$^\prime$] (defined as in Thomas et al. 2003), while the right
panel shows Mgb5177 versus $\langle$Fe$\rangle$ (=$\frac{1}{2}$(Fe5270+Fe5335)). 
In each case, the five pairs of boxes indicate five bins in velocity dispersion; these bins are defined by $\sigma$ ranges
as in Table~8 of Nelan et al., the mean $\sigma$ being indicated in the plot. 
For each bin, the dark-shaded box indicates the 1$\sigma$ error on the weighted mean line-strengths for galaxies with $R_{\rm cl}/R_{200}<0.3$
(approximately the median \rcl), while the light-shaded box shows the same for $0.3<R_{\rm cl}/R_{200}<1.0$. 
Bold vectors show the effect of 15\% increases in age, metallicity and \afe, on each pair of indices, from the models of TMBK.
}
\label{fig:grids}
\end{figure*}

It is interesting to consider ratio the $a_R/a_\sigma$, which compares the effect on a given index of increasing the galaxy mass with 
the effect of increasing the cluster-centric radius. A simple hypothesis would be that galaxies at larger radius have the the same
stellar population properties as more central galaxies of different mass. In this case, the stellar populations would form a one-parameter family; 
for example, this could arise from a single evolutionary process which is switched off at a different redshift, dependent on both $\sigma$ and 
\rcl. 
A one-parameter model of this form would predict that all indices have the same $a_R/a_\sigma$ ratio within the errors. 
In fact, as shown in Table~\ref{tab:fitcoeffs} and Figure~\ref{fig:radsigrat}, there is clearly a variation in this ratio
among the twelve fitted indices. Specifically, in the Balmer indices (HdF, HgF, Hbeta) and the indices
Mgb5177, CN1 and Fe4668, the effect of increasing the radius is equivalent to that of {\it decreasing} the mass. However, 
for most of the Fe indices, the effect of increasing the radius is similar to {\it increasing} the mass. 
We can infer from this that the processes driving the radial systematic behaviour are distinct from those 
which imprinted the \ixsig\ relations themselves\footnote{Moreover, the range in $a_R/a_\sigma$ argues against an explanation 
in terms of an environmentally-driven systematic shift in the velocity dispersion, e.g. a non-universal mass--$\sigma$ relation.}.
This is quite a general conclusion, and does not require any model-dependent
interpretation of the indices. To anticipate the results of the next section however, note that the indices with $a_R/a_\sigma<0$
are those associated with age and \afe\ in the TMBK models, while the indices with  $a_R/a_\sigma>0$ trace mainly the overall 
metallicity \zh. Thus, we may expect that the scaling of the three parameters with cluster-centric radius will differ qualitatively 
from the mass-scaling relations. A similar conclusion can be reached from Figure~\ref{fig:grids}, which shows two classical `two-index'
diagrams. In particular, the diagram of Mgb5177 vs $\langle$Fe$\rangle$ shows that the mass-trend followed by the outer galaxies
is offset from that of the inner objects. The effect of changing cluster-centric radius is not parallel to 
the effect of changing the galaxy mass, again suggesting that at least two distinct mechanisms are at work in determining the average
stellar population at a given mass and radius.

\section{Interpretation of the gradients}\label{sec:interpret}

In this section, we analyse the single-index fit results in terms of the TMBK spectral synthesis models. The models predict the 
values of each index, over a grid in the three-parameter space of age, total metallicity (\zh) and $\alpha$-element to iron-peak
element abundance ratio (\afe). Following the differential method introduced by Nelan et al. (2005), we 
attempt to reproduce the observed index-versus-radius trends (i.e. the values of $a_R$) in terms of underlying radial trends of 
the three model parameters.

\subsection{Parameter responses in the TMBK models}

In principle, given line-strength measurements of sufficiently high signal-to-noise, and assuming the calibration of the
models and data to be secure, it should be possible to convert the index measurements for a single galaxy into 
determinations of age, \zh\ and \afe\ for that object. This `grid-inversion' approach has traditionally utilised a small number of indices
such as Mgb5177, Hbeta and Fe5270/Fe5335 (e.g. Trager et al. 2000; Kuntschner et al. 2001, 2002; Poggianti et al. 2001; Thomas et al. 2005), but can be readily 
extended to a larger index set (Proctor \& Samson 2002; Proctor et al. 2004). 

In practice however, the observed line-strengths are subject to large random errors, so that in general some measurements scatter outside of the
model grid, and because the model grids are tilted with respect to the observed quantities, errors in the derived population parameters are 
correlated, complicating their interpretation (Kuntschner et al. 2001; Thomas et al. 2005). Beyond the problems associated with random errors, 
which can be addressed by considering composite spectra (Bernardi et al. 2006) or averaged indices (Nelan et al. 2005), there are also systematic 
calibration uncertainties in both the data and the models. This is often expressed by the observation that, for instance, the 
overall position of the classical Mgb5177 versus Hbeta diagnostic grid is less secure than the relative differences in predictions. 
These considerations motivated us to introduce a differential analysis of the index--$\sigma$ relations, (Nelan et al. 2005). 
Here, we recap this approach, and then extend the method to describe the cluster-centric trends in the same manner. 

We use the TMBK models to predict the linear `responses' of each index to each of the 
driving parameters, while holding the others constant. Thus for Mgb5177, we determine the following responses:
\[
R^{\rm Mgb5177}_{\rm age} =  \frac{\partial\ \rm Mgb5177}{\partial\ \log {\rm Age}} \ \ ,
\]
\[
R^{\rm Mgb5177}_{{\rm Z}/{\rm H}} = \frac{\partial\ \rm Mgb5177}{\partial\ [{\rm Z}/{\rm H}]} \ \ {\rm  and} \ \
\]
\[
R^{\rm Mgb5177}_{\alpha/{\rm Fe}} = \frac{\partial\ \rm Mgb5177}{\partial\ [\alpha/{\rm Fe}]} \ \ .
\]
Such a description is valid where the model grids are approximately parallel and of constant gradient. This is the chief simplifying assumption, 
and also the main limitation of our method. The responses are computed relative to a fiducial population, and over a subset of the 
TMBK parameter space, roughly corresponding to old, metal-rich, $\alpha$-enhanced ellipticals (Table~\ref{tab:respgrid}). 
For example, we determine $R^{\rm Mgb5177}_{{\rm Z}/{\rm H}}$ over the range $0.00-0.35$ in \zh, holding the age at 8\,Gyr and \afe\
at 0.2. (Strictly, as in Nelan et al., the responses for age and \afe\ are an average of those for fiducial metallicities \zh=0 and those for \zh=0.35.) 

\begin{table}
\caption{Fiducial model and the parameter ranges used to compute parameter responses from the TMBK stellar population models.}
\label{tab:respgrid}
\begin{center}
\begin{tabular}{lcc}
\hline
Parameter & fiducial & range \\
\hline
Age (Gyr)     & 8.0 & \ \ \ \,3.0 $-$ 15.0 \\
$[{\rm Z}/{\rm H}]$         & 0.00,0.35 & $-$0.33 $-$ 0.67 \\
\afe & 0.20 & \ \ \,0.00 $-$ 0.50 \\
\hline
\end{tabular}
\end{center}
\end{table}

\begin{figure}
\includegraphics[width=85mm, angle=270]{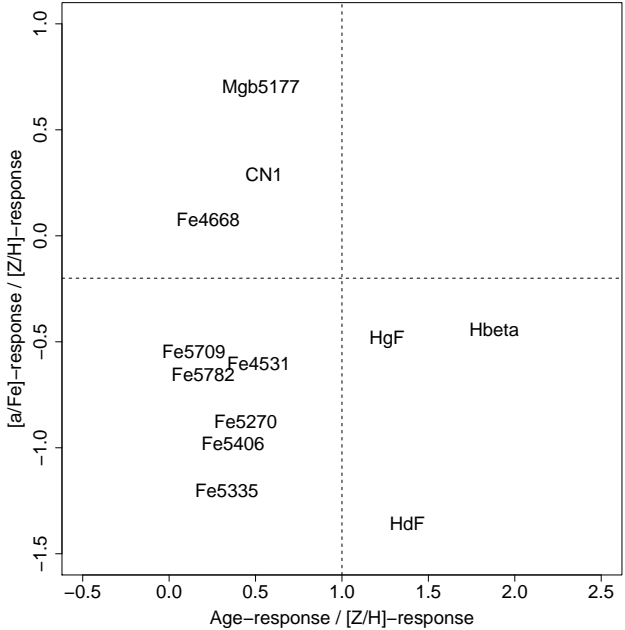}
\caption{Linear responses of indices in the TMBK models, over the parameter range appropriate to early-type galaxies. The age and $\alpha$-abundance response are each normalized by the metallicity response, to reduce the information to a two-dimensional format. The quadrants, formed by arbitrarily-placed vertical and horizontal lines, divide the indices according to the parameters which they best constrain, as in Figure~\ref{fig:radsigrat}.}
\label{fig:responses}
\end{figure}

The complete set of response vectors, as used for the analyses described below, is summarized in Table~\ref{tab:responses}. 
The table also indicates the scatter, $S$, in the response, computed over all pairs of TMBK models within the range specified in 
Table~\ref{tab:respgrid}. This quantity serves as a conservative estimate of the error introduced by forcing a linear
description of the grids. In Figure~\ref{fig:responses}, we show the age and \afe\ responses of the twelve indices
relative to their \zh\ response. This figure neatly separates the three classes of index described in Section~\ref{sec:data}, 
and in particular shows why we classify Fe4668 with Mgb5177 and CN1 rather than with the other Fe indices. It also shows why the behaviour
of HdF is expected to differ from that of Hbeta and HgF, as discussed later. 

\begin{table}
\caption{Index responses to population parameters. $R_{\rm par}$ is the change in the index per decade in parameter `par', estimated 
over the range given in Table~\ref{tab:respgrid}, and holding the other parameters at their fiducial values. As an estimate
of the errors in the responses, $S_{\rm par}$ is the rms scatter in $R_{\rm par}$ as computed over pairs of models within 
the parameter range. The quantities are for the line-strengths expressed in Angstroms, except for CN1 in magnitudes. The
divisions in the table reflect the relative responses, as in Figure~\ref{fig:responses}.}
\label{tab:responses}
\begin{tabular}{lrrrrrr}
\hline
Index      &
$R_{\rm age}$ &
$R_{\rm Z/H}$ &
$R_{\alpha/{\rm Fe}}$ & 
$S_{\rm age}$ &
$S_{\rm Z/H}$ &
$S_{\alpha/{\rm Fe}}$ \\
\hline
HdF     &--1.921& --1.393&  1.889 &  0.445 &  0.414 &  0.752 \\
HgF     &--2.803& --2.223&  1.091 &  0.599 &  0.391 &  0.214 \\
Hbeta   &--1.134& --0.603&  0.266 &  0.211 &  0.189 &  0.053 \\
\hline
CN1     & 0.098&  0.177&  0.052 &  0.046 &  0.032 &  0.019 \\
Mgb5177 & 1.547&  2.914&  2.026 &  0.349 &  0.498 &  0.634 \\
Fe4668  & 1.748&  7.720&  0.608 &  0.918 &  0.740 &  0.707 \\
\hline
Fe4531  & 0.812&  1.565& --0.940 &  0.243 &  0.223 &  0.178 \\
Fe5270  & 0.660&  1.488& --1.305 &  0.242 &  0.175 &  0.230 \\
Fe5335  & 0.577&  1.729& --2.078 &  0.179 &  0.461 &  0.598 \\
Fe5406  & 0.437&  1.181& --1.157 &  0.145 &  0.208 &  0.237 \\
Fe5709  & 0.082&  0.573& --0.312 &  0.048 &  0.033 &  0.019 \\
Fe5782  & 0.104&  0.527& --0.344 &  0.064 &  0.036 &  0.027 \\
\hline
\end{tabular}
\end{table}

As in Nelan et al., we model the observed \ixsig\ slopes in terms of underlying linear trends of log(age), \zh\ and \afe, as a function of $\log\sigma$, 
using the response vectors to translate between the observed and underlying relationships:
\[
a_\sigma = \frac{d I}{d \log\sigma} =
\]
\[
	\frac{\partial I}{\partial \log {\rm age}} \cdot \frac{\partial \log {\rm age}}{\partial \log\sigma} +
	\frac{\partial I}{\partial [{\rm Z}/{\rm H}]} \cdot \frac{\partial [{\rm Z}/{\rm H}]}{\partial \log\sigma} +
	\frac{\partial I}{\partial [\alpha/{\rm Fe}]} \cdot \frac{\partial [\alpha/{\rm Fe}]}{\partial \log\sigma}
\]
\begin{equation}
= R^{index}_{\log {\rm age}} P^{\log {\rm age}}_\sigma + R^{index}_{[{\rm Z}/{\rm H}]} P^{[{\rm Z}/{\rm H}]}_\sigma + R^{index}_{[\alpha/{\rm Fe}]}  P^{[\alpha/{\rm Fe}]}_\sigma
\end{equation}
where the change of each parameter with $\log\sigma$ has been written as $P^{\rm par}_\sigma$ for brevity.

We can extend the Nelan et al. method to the radial trends, by writing an equivalent expression to describe the $a_R$ coefficients in 
terms of underlying changes of age, metallicity and $\alpha$-abundance as a function of cluster-centric radius, $R_{\rm cl}$:

\begin{equation}
a_R = R^{index}_{\log {\rm age}} P^{\log {\rm age}}_{R_{\rm cl}} + R^{index}_{[{\rm Z}/{\rm H}]} P^{[{\rm Z}/{\rm H}]}_{R_{\rm cl}} + R^{index}_{[\alpha/{\rm Fe}]}  P^{[\alpha/{\rm Fe}]}_{R_{\rm cl}}
\end{equation}

With at least three indices having different response signatures relative to the three population parameters, we can attempt to invert these equations 
for the elements of $P_\sigma$ and $P_{R_{\rm cl}}$, and hence determine how age, metallicity and $\alpha$-abundance scale with mass and with location 
in the cluster. A minimal set of indices for this purpose might be HgF, Mgb5177 and Fe5270. The first of these provides age sensitivity while the remaining 
two trace metallicity, with differing dependency on $[\alpha/{\rm Fe}]$. In what follows, we use not only these three but all of the twelve indices chosen 
in Section~\ref{sec:data}, to over-constrain the solution. We cast the fit in terms of predicting the coefficients $a_R$, with the free parameters 
$P^{\rm par}_{R_{\rm cl}}$ to be determined. In the fitting stage, we assume the $R^{index}_{\rm par}$ are known precisely, and weight the data
only according to the errors on the measured $a_R$. 

In summary, we emphasize that our strategy for determining the parameter gradients $P$ is explicitly a two-step process: first, we fit the 
line-strength data for the $\sim$3000 galaxies to determine the radial coefficients of the indices; second, we model these twelve coefficients $a_R$ by 
fitting to the model responses, reducing the information to just three parameters. If a good fit can be obtained, the slopes of age, \zh\ and \afe\ with radius
provide an economical description of the observed trends. A weakness of this approach is that it implicitly assumes the errors on the $a_R$ are 
not correlated with each other, such that each index slope constitutes an independent constraint on the model. A more sophisticated approach would attempt
to combine the two steps into a single fit. The present data probably do not justify such a treatment, and we prefer to retain the intermediate step 
for clarity of presentation. 

\subsection{Stellar population trends along the $\sigma$ sequence}

\begin{figure*}
\includegraphics[width=57mm, angle=270]{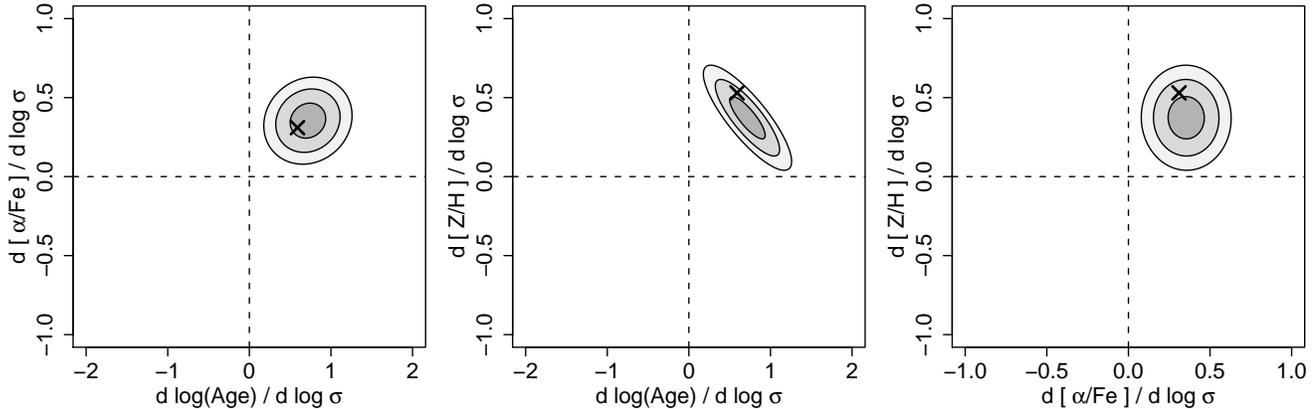}
\caption{Constraints on variation of TMBK population parameters with velocity dispersion. The ellipse boundaries indicate the 68\%, 95\% and 99\% confidence levels in the two parameters taken jointly. The black crosses show the results derived by Nelan et al. (2005).}
\label{fig:sigellipses}
\end{figure*}

First, we revisit the mass scaling relations, as earlier discussed by Nelan et al. (2005). Figure~\ref{fig:sigellipses} shows the 
confidence intervals on the quantities $P^{\rm par}_\sigma$, derived from the \ixsig\ slopes $a_\sigma$, as in Table~\ref{tab:fitcoeffs}
We derive the following scalings: 
\begin{equation}\label{eqn:sigmod}
\rm{Age}\propto\sigma^{0.72\pm0.14} , \  \rm{Z/H}\propto\sigma^{0.37\pm0.08} , \  \alpha/{\rm Fe}\propto\sigma^{0.35\pm0.07}, 
\end{equation}
confirming significant trends of all three parameters with $\sigma$, so that on increasing the velocity dispersion, the stellar
populations become older, more metal rich, and more $\alpha$-enhanced. 
As indicated in the figure, there is a strong degeneracy between the age and metallicity scalings, such that the errors on 
these parameters are correlated. 
Comparison with the trends reported by Nelan et al. shows that despite several differences in the sample construction
and analysis methods (notably, the presence of additional terms in the \ixsig\ model), the $a_\sigma$ coefficients
yield scaling relations consistent with the previous results, at the $\sim$1$\sigma$ level. 

Figure~\ref{fig:sigmodsbest} provides a visual impression of the success of Equation~\ref{eqn:sigmod} in reproducing the measured \ixsig\ slopes 
(cf. Fig. 10a of Nelan et al.). 
We plot the {\it measured} slopes for each index on the horizontal axis, normalized by the measurement error. Thus, the 
position of each index on the horizontal axis reflects its significance, and the horizontal errors are unity by definition. 
The vertical axis gives the {\it predicted} index--$\sigma$ slopes, under the scalings of Equation~\ref{eqn:sigmod}, also normalized by
the error in the {\it measurement}. 
The errors in the predictions are derived from the scatter in the responses over the relevant subregion of the model grid, i.e. the $S_{\rm par}$
given in Table~\ref{tab:responses}\footnote{
For the mass-scalings discussed here, errors in the predictions dominate over the errors in the observed slopes. However, the 
prediction errors are only indicative, and are not independent from index to index. Thus we do not attempt to derive a formal goodness-of-fit
in this case. In contrast, for the radial trends analysed later, the observed errors are comparable to the prediction errors, and a $\chi^2$ can 
be quoted with greater justification.}. For an adequate solution, the error boxes for all indices would lie along the line of unit slope. Outliers
from this line highlight indices with $a_\sigma$ coefficients not well described by the model. For instance, as discussed by Nelan et al., Fe5709
decreases with increasing $\sigma$, although our model would predict it to increase, driven primarily by \zh.

\subsection{Cluster-centric trends in stellar populations}\label{sec:radparvar}

We now turn to the main objective of this paper, which is to determine which aspects of the stellar populations are different in the 
cores of clusters compared to the outer regions, and are responsible for driving the radial dependencies of the \ixsig\ residuals. 
Unlike many previous works (Guzm\'an et al. 1992; Carter et al. 2002), our approach explicitly allows for simultaneous variations
in age, total metallicity and $\alpha$-abundance ratio. 

Figure~\ref{fig:radellipses} shows the confidence intervals on the quantities $P^{\rm par}_{R_{\rm cl}}$, derived from the coefficients $a_R$. 
The key result is that both age and \afe\ are found to depend on cluster-dentric distance, while no there is no significant trend of \zh: 
\[
\frac{{\rm d}\log{}{\rm Age}}{{\rm d}R_{\rm cl}} \ = -0.072\pm0.017 
\]
\begin{equation}\label{eqn:radmod}
\frac{{\rm d}[{\rm Z/H}]}{{\rm d}R_{\rm cl}} \ \ \ \ = +0.007\pm0.010 
\end{equation}
\[
\frac{{\rm d}[\alpha/{\rm Fe}]}{{\rm d}R_{\rm cl}} \ \ \ = +0.048\pm0.009 
\]
Thus, galaxies at 1\,$R_{200}$ are on average 15\% younger and 10\% less $\alpha$-enriched than galaxies of the same mass located in the core of 
the cluster. Considering the one-parameter confidence intervals, the significance of the age and \afe\ variations are above the 4$\sigma$ and 
5$\sigma$ levels, respectively.

\begin{figure}
\includegraphics[width=85mm,angle=270]{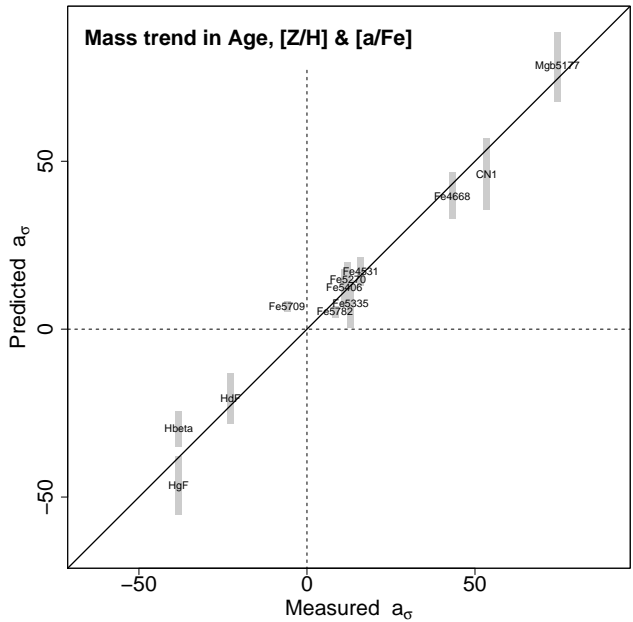}
\caption{Predicted versus observed index--$\sigma$ slopes in the best fitting model for the trends with mass. 
Both the model predictions and the observed slopes are scaled by the error in the observed slope. The shaded box shows the $1\sigma$ errors, 
where the errors in the predictions are derived from the errors in the responses (Table~\ref{tab:responses}). Indices which deviate significantly 
from the unit-slope line have slopes poorly reproduced by the model.}
\label{fig:sigmodsbest}
\end{figure}

\begin{figure*}
\includegraphics[width=57mm, angle=270]{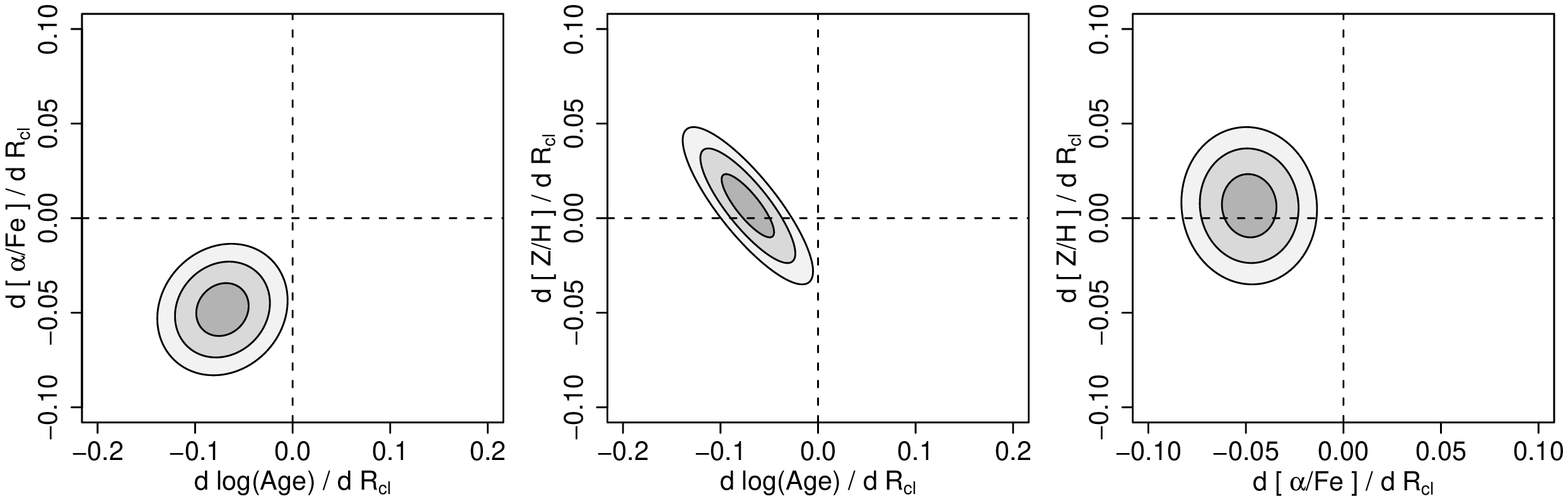}
\caption{Constraints on variation of TMBK population parameters with cluster-centric radius. The ellipse boundaries indicate the 68\%, 95\% and 99\% confidence levels in the two parameters taken jointly.}
\label{fig:radellipses}
\end{figure*}

\begin{figure}
\includegraphics[width=85mm, angle=270]{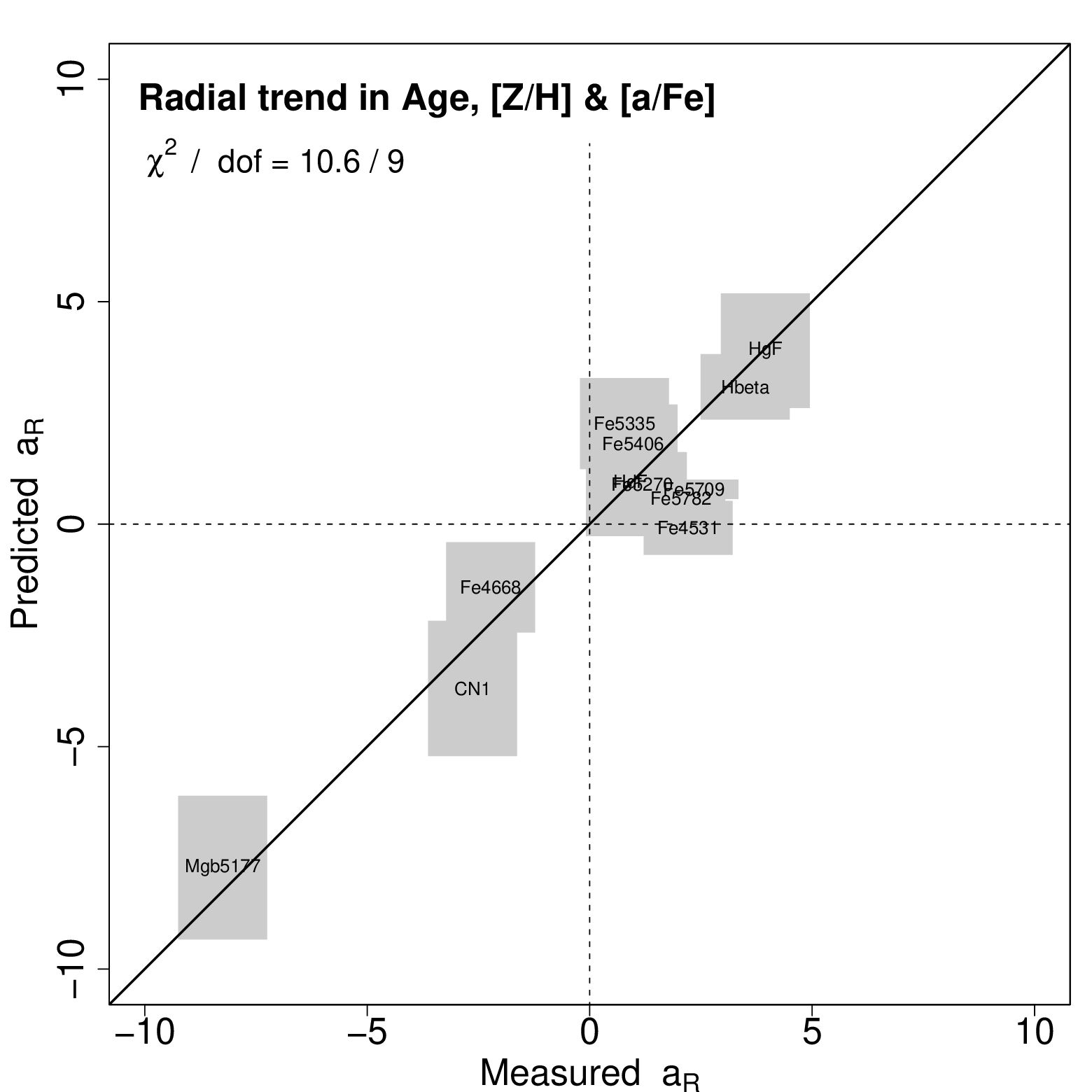}
\caption{Predicted versus observed slopes in the best fitting model for the trends with cluster-centric radius. The diagram
is equivalent to Figure~\ref{fig:sigmodsbest} for the mass-scalings.}
\label{fig:radmodsbest}
\end{figure}

The success of this model, in reproducing the observed cluster-centric trends in the indices, can be judged from Figure~\ref{fig:radmodsbest}, 
which compares the measured and predicted $a_R$ for each index. The figure confirms that the model indeed accounts for the decrease of Mgb5177, 
CN1 and Fe4668, and also the increase of Hbeta and HgF with increasing radius. The case of the remaining Balmer index HdF is instructive. 
The age-sensitivity of HdF would lead to a positive slope with radius, if the age-radius trend acted alone. However, compared to Hbeta and HgF, 
the HdF index also responds strongly to \afe, as seen in Figure~\ref{fig:responses}. 
At large radius, the HdF is strengthened by the younger ages, but weakened by the lower \afe. Our model
predicts the effects of age and \afe\ gradients will compensate each other in the case of HdF, successfully 
reproducing the observed fairly flat behaviour. The Fe-sensitive indices are predicted to have weak trends, as observed. 
The model predictions for each index are also over-plotted on the partial residual diagrams in Figure~\ref{fig:partreds}, for comparison to the 
linear fits for $a_R$, and to the binned data. It is seen that only in the case of Fe4531 is the model prediction outside of the 2$\sigma$ 
observational limits. In contrast to the equivalent diagram for the mass-scalings (Figure~\ref{fig:sigmodsbest}), 
the measurement errors are comparable to the errors in the predictions. Taking both into account, the model has 
$\chi^2=10.6$ for nine degrees of freedom (twelve parameters less three fitted coefficients), and thus is formally an acceptable fit. 
Hereafter, we refer to the solution given in Equation~\ref{eqn:radmod} as the `default model'; its parameters are summarized in 
Line~1 of Table~\ref{tab:rcltrends} for reference. For comparison,~Line 2 of the table gives the $\chi^2=126$ obtained for the null model 
(no dependence of stellar populations on cluster-centric radius).

\begin{table*}
\caption{Cluster-centric trends of age, \zh\ and \afe. The various lines record the results of tests discussed in the text, as follows:
Line 1 gives the results from the default three-parameter model (Section~\ref{sec:radparvar}); 
Line 2 gives the $\chi^2$ for the null hypothesis for comparison;
Lines 3--5 show the best-fitting two-parameter models;
Lines 6--8 show the best-fitting one-parameter models; 
Lines 9--10 investigate the influence of morphology on the solution (Section~\ref{sec:morpheff}); 
Lines 11--21 test for systematic errors by fitting subset of the galaxy sample (Section~\ref{sec:subsample}); 
}
\label{tab:rcltrends}
\begin{tabular}{clccccl}
\hline
&    & log(Age) gradient & \zh\ gradient & \afe\ gradient & $\chi^2$/dof & notes \\
\hline
1&Default         & $-0.072\pm0.017$ & $+0.007\pm0.010$ & $-0.048\pm0.009$ & \z10.6 / \z9 & default solution (Equation~\ref{eqn:radmod}) \\ %updated
\hline
2&No gradients    &  --- & --- & --- &  126.1 / 12 & null model \\ %updated
\hline
3&age \& \afe     &  $-0.063\pm0.009$ & ---              & $-0.048\pm0.008$ & \z12.0 / 10 & no radial \zh\ trend \\ %updated
4&\zh\ \& \afe    &  ---              & $-0.031\pm0.010$ & $-0.043\pm0.014$ & \z40.1 / 10 & no radial age trend \\ %updated
5&age \& \zh      &  $-0.059\pm0.033$ & $+0.004\pm0.021$ & ---              & \z50.0 / 10 & no radial \afe\ trend \\ %updated
\hline
6&Age only        & $-0.054\pm0.017$  & ---              & ---              & \z53.1 / 11 & no \zh\ or \afe\ trends \\ %updated
7&\afe\ only      &  ---              & ---              & $-0.037\pm0.019$ & \z75.5 / 11 & no age or \zh\ trends \\ %updated
8&\zh\ only       &  ---              & $-0.027\pm0.012$ & ---              & \z79.7 / 11& no age or \afe\ trends \\ %updated
\hline
9&$B/T>0.5$       & $-0.040\pm0.023$  & $-0.013\pm0.014$ & $-0.048\pm0.012$ & \z\z8.9 / \z9 & bulge-dominated \\ %updated
10&fit $B/T$       & $-0.066\pm0.024$  & $+0.002\pm0.015$ & $-0.046\pm0.013$ & \z12.5 / \z9 & control for $B/T$ in index fits\\ %updated
\hline
11&$\log\sigma>2.10$& $-0.051\pm0.013$  & $-0.001\pm0.008$ & $-0.054\pm0.007$ & \z\z5.9 / \z9 & excludes $\sim\frac{1}{3}$ of galaxies\\ % updated
12&$\log\sigma<2.24$& $-0.096\pm0.019$  & $+0.004\pm0.012$ & $-0.048\pm0.010$ & \z\z7.1 / \z9 & excludes $\sim\frac{1}{3}$ of galaxies\\ % updated
13&DX clusters      & $-0.100\pm0.021$  & $+0.018\pm0.013$ & $-0.051\pm0.011$ & \z\z9.7 / \z9 & coincident cD and X-ray peak\\ % updated
14&$cz<13800$\,\kms& $-0.061\pm0.029$  & $+0.014\pm0.018$ & $-0.045\pm0.014$ & \z14.1 / \z9 & clusters within median depth\\ % updated
15&$cz>13800$\,\kms& $-0.078\pm0.033$  & $-0.001\pm0.021$ & $-0.047\pm0.018$ & \z22.8 / \z8 & clusters beyond median depth\\ % updated
16&Dec. $>0^\circ$ & $-0.096\pm0.026$  & $+0.015\pm0.015$ & $-0.035\pm0.013$ & \z10.3 / \z9 & mostly WIYN\\ % updated
17&Dec. $<0^\circ$ & $-0.062\pm0.014$  & $+0.009\pm0.009$ & $-0.068\pm0.008$ & \z\z4.5 / \z9 & mostly CTIO\\ % updated
18&No $M_J$ cut    & $-0.081\pm0.015$  & $+0.011\pm0.009$ & $-0.051\pm0.008$ & \z\z9.2 / \z9 & includes cDs \\ % updated
19&No emission cut  & $-0.073\pm0.017$  & $+0.003\pm0.010$ & $-0.051\pm0.008$ & \z\z9.5 / \z9 & No cut in \oiii\ or \hb \\ % updated
20&\rcl$<2.0R_{200}$&$-0.068\pm0.017$  & $+0.007\pm0.010$ & $-0.033\pm0.009$ & \z16.7 / \z9 & increased radial range of fit\\ % updated
21&\rcl$<0.5R_{200}$&$-0.139\pm0.030$  & $+0.028\pm0.019$ & $-0.072\pm0.015$ & \z13.1 / \z9 & reduced radial range of fit\\
\hline
\end{tabular}
\end{table*}

\begin{figure*}
\includegraphics[width=220mm, angle=270]{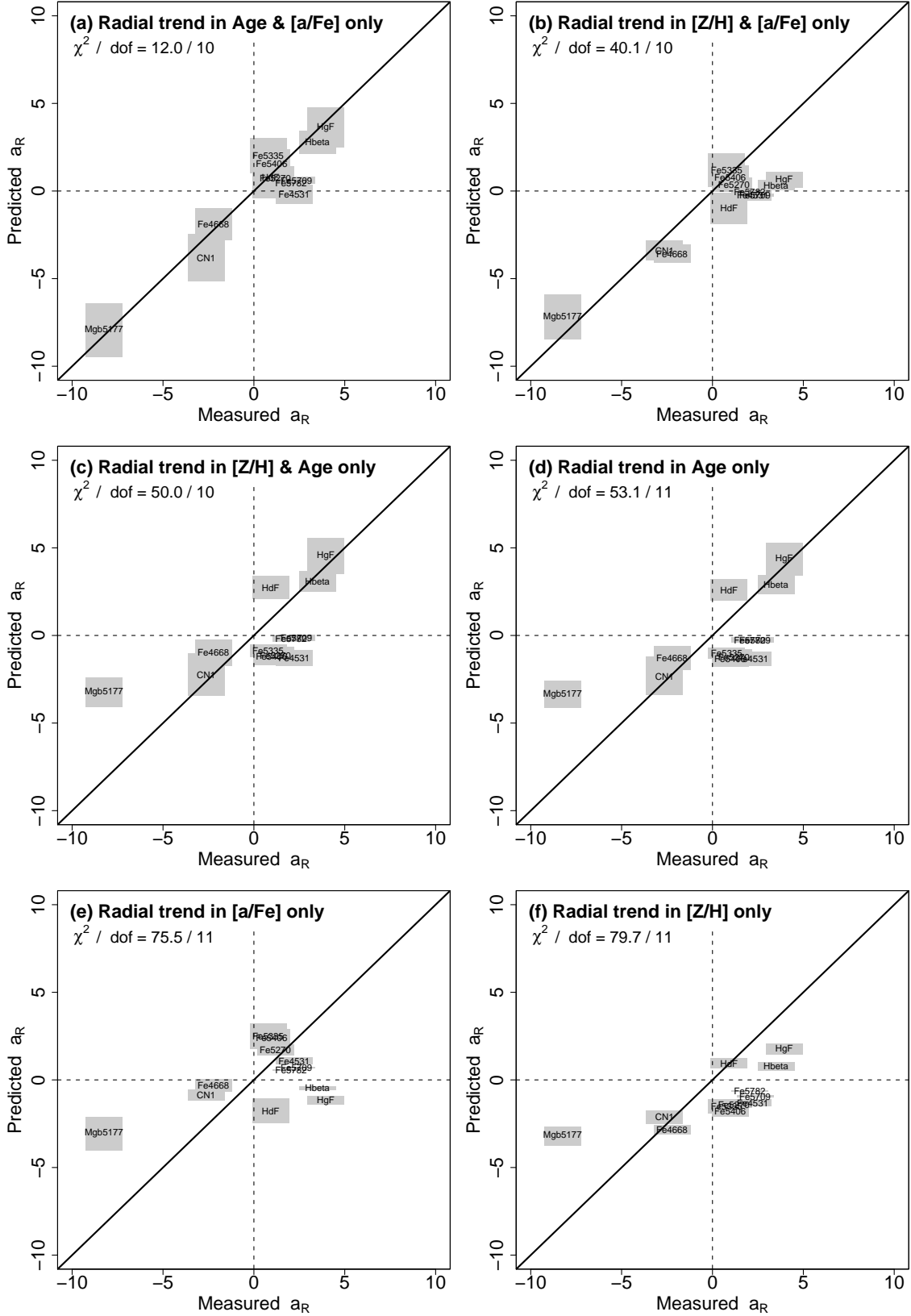}
\caption{Predicted versus observed radial slopes for the restricted models, i.e. where only one or two of the stellar population
parameters are allowed to vary with radius. Each panel is equivalent to Figure~\ref{fig:radmodsbest}, but for the best-fitting
models of the class indicated.}
\label{fig:radmods}
\end{figure*}

To illustrate further the influence of particular indices in our preferred solution, 
we consider whether the $a_R$ can be adequately fit by `restricted' models, in which only one or two of the stellar population parameters are 
allowed to vary with \rcl. The results are shown in Figure~\ref{fig:radmods}, where each panel is equivalent to Figure~\ref{fig:radmodsbest}, 
but now for the best-fitting restricted model of the given type. It is important to note that this is not a projection of the default fit 
into a reduced parameter-space; i.e. the \zh-vs-radius slope of the model considered in Figure~\ref{fig:radmods}d (for instance) is not the same 
as that in Figure~\ref{fig:radmodsbest}. 
As expected from the coefficients of the default model, removing the radial dependence of \zh\ does not significantly compromise the fit 
(Figure~\ref{fig:radmods}a and Line 3 of Table~\ref{tab:rcltrends}). 
By contrast, we find a very poor fit for all of the models which exclude either the age or \afe\ variations.
In particular, the best-fitting model without cluster-centric age variations (Figure~\ref{fig:radmodsbest}b and Line 4)
cannot reproduce the increase of Hbeta and HgF with radius. 
Models without cluster-centric \afe\ variations (Figure~\ref{fig:radmods}c and Line 5) cannot account for the strong negative 
trend of Mgb5177, and also predict a weak negative trend for the Fe indices, rather than the positive trends observed. 
Finally, we obtain extremely poor fits for models which attempt to account for the radial trends in terms of a single driving 
parameter (Figure~\ref{fig:radmods}d,e,f and Lines 6--8). Taking these results at face value, any single-parameter model, 
e.g. a pure age trend (as in Guzm\'an et al. 1992) or a pure metallicity trend (as in Carter et al. 2002), is bound to yield 
inconsistent results when applied to multiple indices.

\subsection{Influence of morphological trends}\label{sec:morpheff}

A very pertinent concern is whether the observed gradients truly represent changes in stellar populations of otherwise similar 
galaxies, or whether instead they reflect a radial variation of the morphological mixture in clusters. 
Recall that the galaxy sample was selected using colour and magnitude information only, without reference to galaxy morphology
(Smith et al. 2004). Additionally, for the present analysis, we imposed a selection against objects with nebular emission lines in the (central)
spectra. Thus although star-forming discs should be excluded, is is very likely that our sample includes some bulges of 
early-type spirals. If such systems have different stellar populations from pure ellipticals of the same mass, and if
the fraction of spirals in the sample increases in the cluster outskirts, then this might be manifest as radial trends
of the kind observed. 

At present, quantitative morphological information is available 
for approximately half of the NFPS galaxies, and of these around 70\% have bulge-dominated morphologies (bulge-to-total 
light ratios, $B/T$, greater than 0.5). One simple test for morphology-driven trends is to fit only these bulge-dominated galaxies. 
In this case, the detection of an age trend is reduced to a confidence level below 2$\sigma$, although the \afe\ trend remains 
significant (Line 9 of Table~\ref{tab:rcltrends}). However, the errors are large after this sub-sampling, 
and the derived gradients are also less than 2$\sigma$ away from our default solution. If the difference is real, 
this might mean that part of the radial trends are driven by changing morphological composition with radius. However, 
it could also result from a true radial age effect (i.e. at fixed morphology), but due to a mechanism that is more 
effective in galaxies with discs. 

An alternative way to test for morphology effects is to allow an additional term proportional to $B/T$ in the \ixsig\ fits. 
To first order, this term should absorb any systematic dependence on morphology, so that the \rcl\ term describes
the radial effect at fixed bulge-fraction as well as fixed mass. Fitting this more complex model to the subsample with 
measured $B/T$ indeed reveals a strong morphology dependence in the index data. The pattern of effects 
is different again from either the mass sequence or the radial trends. Specifically, the Balmer lines and Fe-dominated 
indices are all decreased for more bulge-dominated systems, while the $\alpha$-dominated indices Mgb5177 and CN1 increase 
with greater bulge-fraction. We can quantify this effect in terms of stellar population parameters, by modelling the morphology
coefficients $a_{B/T}$ in the same way as we treated $a_\sigma$ and $a_R$. 
We find that for a shift of one unit in $B/T$ (i.e. going from pure disc to pure bulge), we 
increase log(Age) by $0.176\pm0.026$, 
decrease \zh\     by $0.097\pm0.016$ and
increase \afe\    by $0.052\pm0.013$. 
In summary, galaxies with larger bulges are older, less metal rich and more $\alpha$-enhanced than diskier objects of 
the same velocity dispersion. At face value, this is dissimilar from the conclusions of Thomas \& Davies (2006), who
find that early-type spiral bulges have similar stellar populations to ellipticals of the same velocity dispersion. 

Having controlled for the morphology effects, we can now assess their impact on the cluster-centric trends.
Perhaps surprisingly, the corrected radial trends 
(Table~\ref{tab:rcltrends}, Line 10)
are little changed relative to our default solution.
This robustness arises because the morphology--radius relation is rather shallow for our sample, which was already selected
to have red colours and no nuclear emission. From the core of the cluster out to the virial radius, the average $B/T$ is
reduced only by $0.073\pm0.014$. Combining this figure with the above morphology effects yields a `correction' 
which is consistent with the difference between the default and the morphology-controlled solutions. 
In passing we note that the mass-scaling trends are also not strongly affected by the inclusion of the $B/T$ term, despite a strong
morphology-$\sigma$ relation ($B/T$ increasing by $0.39\pm0.03$ per decade in $\sigma$). In this case, 
when multiplied by the morphological effect on the stellar populations above, the net effect is an order of magnitude 
smaller than the mass trend at constant morphology. For instance, for one decade in $\sigma$, morphology drives
an age change of $0.18\times0.39=0.07$ in the log, compared to the derived relation Age\,$\propto\sigma^{\sim0.7}$.

We conclude that although red-sequence galaxies of different morphologies indeed seem to have systematically different stellar 
populations, this has little impact on the results derived here for the cluster-centric trends, or for the mass-scalings.
An investigation of morphology effects on the NFPS colour--magnitude
relations is presented by Stevenson et al. (2006). A full synthesis of morphological and stellar population 
information must await completion of the $B/T$ measurements, and will be addressed in a future paper.

\subsection{Further robustness tests and data subsamples}\label{sec:subsample}

In this section we continue to address the robustness of the results, by testing the fits obtained after
sub-sampling the input data set. The results are summarized in the lower section of Table~\ref{tab:rcltrends}. 

First, motivated by the apparent mass-dependence of the HgF trends seen in Figure~\ref{fig:quantrun}, 
we test whether the results are strongly driven by the low-mass objects.
Restricting the sample to galaxies with fairly high velocity dispersion $\sigma>125$\,\kms\ removes around one third of 
the objects. Fitting to the reduced sample yields a slightly reduced age gradient, but the solution remains marginally consistent
with our default fit (Line~11 of Table~\ref{tab:rcltrends}). Similarly, restricting to $\sigma<175$\,\kms\, yields a marginally
stronger radial age gradient (Line~12); the \afe\ and \zh\ trends are unaffected by cuts in $\sigma$. Despite sharing 50\% of galaxies
in the intermediate mass range, the two subsamples differ from each other at $\sim$2$\sigma$ in the recovered age gradient. 
Thus there is some evidence that the age trends are more pronounced for galaxies of smaller mass. Because of the tendency towards more 
disc-dominated objects at low $\sigma$, this test is not independent of the $B/T$ cut in Section~\ref{sec:morpheff}. Note also that 
because the average stellar populations vary strongly with mass, we cannot make arbitrarily severe cuts on $\sigma$ without 
potentially invalidating the response vectors adopted. 

A potential concern is that some clusters have ambiguous or poorly-determined centres (Smith et al. 2004), such that, in the worst cases, the 
\rcl\ values used in the fits will not be a good proxy for galaxy environment. If we use only the galaxies in clusters
with the most secure centres (`DX', based on a dominant galaxy coincident with an X-ray emission peak), we obtain a marginally 
stronger and more significant age trend (Line~13). This suggests that clusters with less well-determined centres are diluting
the trends in the default solution, either because the radii are less meaningful or else because more relaxed clusters have intrinsically
stronger gradients. In this context, we note that in some well-sampled systems such as instance Abell 3558, the radial gradients
in Mgb5177, etc, can be detected for individual clusters. This in itself confirms that the results obtained for the full sample
are not simply an artefact of stacking the clusters. 

In more distant clusters, we typically sample more luminous galaxies on average, and to somewhat larger cluster-centric radius. 
Moreover, any instrumental systematic error affecting data over a particular wavelength range will have an impact on
different indices at different redshifts. A simple test for such effects is to divide the sample into halves according to 
host cluster redshift, as in Lines~14--15. The two cluster sets yield radial trends which are consistent with each other, 
and with the default solution, within the enlarged errors. 
An alternative division of the cluster set is into northern and southern celestial hemisphere, which 
are dominated, respectively, by data obtained at the WIYN and CTIO telescopes. Systematic offsets between 
index measurements at one telescope relative to another have been investigated at the data level by Nelan et al. (2005). 
Any residual offsets could in principle lead to radial trends in the data since the WIYN field of view 
was wider than the equivalent field of view at CTIO. Fitting the northern and southern cluster sets separately, we find a 
weaker \afe\ gradient in the northern data than the southern data (at the $\sim$2$\sigma$ level; Lines~16--17). Each is marginally
inconsistent with the default solution. The age gradients are consistent between the subsamples.

Next we relax some of the selection criteria used to define the galaxy sample, to check how these affect the 
results. Removing the cut on J-band luminosity, allowing the cDs (and other large galaxies) to re-enter into the fits, 
there is no significant effect on the results (Line~18). If we allow galaxies with nebular emission to contribute 
to the fits, the solution is again almost unaffected (Line~19). This surprising stability arises partly because the iterative clipping
process efficiently removes many of the galaxies with emission, and partly because we are using multiple Balmer indices, 
with the more robust HgF and HdF balancing the more easily contaminated Hbeta. 

Finally we test the effect of changing the outer radial cut. 
We noted in Section~\ref{sec:fitmethod} that a few clusters have unreliable $R_{200}$ due to poorly-determined cluster
velocity dispersions. In such clusters, the \rcl/$R_{200}$, are subject to systematic errors. 
To reduce the influence of galaxies in these clusters, 
we limited the fit to \rcl$<R_{200}$. Replacing some of these galaxies, and fitting now over a larger range, 
\rcl$<2R_{200}$, we derive a consistent age trend, while the trend of \afe\ is marginally weakened (Line~20). 
If instead we restrict the range still further, to \rcl$<0.5R_{200}$, both the age and \afe\ trends become
stronger (by $\sim$2$\sigma$ and $\sim$1.5$\sigma$ respectively), albeit with errors 50\% larger (Line~21). This behaviour
can be discerned in the radial residuals diagram (Figure~\ref{fig:partreds}), where a number of the indices
show a stronger trend in the inner few bins, and are flatter further out. 
Motivated by the apparently stronger gradients in the inner regions, we have experimented with an alternative parametrization of the 
radial trends, using a term proportional to $\log(R_{\rm cl}/R_{200})$. This requires an ad hoc treatment of the galaxies at
very small \rcl, to prevent them from exercising excessive leverage on the fits. If we place the inner galaxies
at a minimum radius of 0.01\,$R_{200}$, we recover an age trend of 12$\pm$2\% per decade in $R_{\rm cl}/R_{200}$ and 
an $\alpha$/Fe trend of 7$\pm$1\% per decade.  These results are, however, sensitive the treatment adopted for the 
innermost galaxies. 

In conclusion, among all of the data subset tests performed above, significant trends towards younger ages and
lower \afe, at larger cluster-centric radius, are recovered in every case. In addition to demonstrating the robustness
of the cluster-centric variations, these tests have hinted at a number of possible departures from the assumptions
of our modelling. First, there is some evidence that the cluster-centric trends are steeper in the cluster core than 
further out; alternative parametrizations of the \rcl\ term could in principle yield improved fits. 
Second, the trends appear to be stronger for low-mass and/or disky galaxies than for high-$\sigma$ bulge-dominated systems. 
Our model (Equation~\ref{eqn:ixsigmod}) does not naturally account for such an effect, since it assumes linear addition of mass and radius contributions, with 
no cross term (i.e. no term proportional to the product $R_{\rm cl}\log\sigma$). However, the results of the robustness tests might be
pointing towards real differences in the impact of the environment on different galaxy types and masses. 

\section{Discussion}\label{sec:discuss}

In this section we first compare our results with previous work on trends in red-sequence galaxy properties 
within clusters. We also review the distinct approach of comparing cluster galaxies with their counterparts
in the low-density field, and its relation to the present study. Finally, we discuss recent predictions
for cluster-centric age gradients from N-body simulations and semi-analytic modelling. 

\subsection{Cluster-centric gradients in stellar populations}

Our present study is, of course, not the first to address the issue of environmental effects on early-type galaxy properties. 
In comparing to previous results, we concentrate on primarily spectroscopic tests, which can more easily separate age and metallicity
effects. Among published studies, 
the closest in spirit to our approach are those of Guzm\'an et al. (1992) and Carter et al. (2002). Each of these works used
line-strengths for galaxies in only a single cluster, Coma, in contrast to our results which implicitly average over many clusters. 

Guzm\'an et al. (1992) reported a zero-point offset of $0.017\pm0.005$\,mag in the Mg$_2-\sigma$ relation between a sample from the core of
Coma and a `halo' sample at larger radius. Their study included only morphologically-classified ellipticals, probing typically typically higher mass systems than our sample (90\% with $\sigma>150$\,\kms, the median value for NFPS). 
The sense of the offset is the same as observed here, i.e.
that at fixed velocity dispersion, the Mg line-strength is 
weaker at larger radius. 
With only one index, Guzm\'an et al. could not distinguish among alternative interpretations of the offset as attempted here, but instead described 
the effect in terms only of age, with more recent star-formation occurring in the outer galaxies. 
The median radius for the two samples corresponds to $0.3\,R_{200}$ and $1.5\,R_{200}$. 
Over this range, the radial term in our fit for the Mgb5177 index would imply a change of 0.18\,\AA, 
equivalent to 0.011\,mag in Mg$_2$. Quantitatively, then, the radial behaviour of Mg found 
in this paper is marginally weaker than that claimed by Guzm\'an et al. 

Carter et al. (2002) investigated residual gradients using
Mg$_2$, Hbeta and $\langle$Fe$\rangle$ indices for a sample extending to 1\,$R_{200}$ (80\,arcmin) in Coma, without morphological 
or colour selection. They report a very significant radial trend in Mg$_2$, with a change of 0.03\,mag from core to $R_{200}$. 
Carter et al. interpret their results as metallicity differences at fixed mass, and comment on the
possibility of an abundance ratio effect, since their measured Fe trend is weak. 
Converting 
to Mgb5177, the implied radial trend is $\sim0.6\pm0.2$\,\AA\ per $R_{200}$. Thus the Carter et al. trend, although in the
same direction as our result, is four times steeper (and three times steeper than Guzm\'an et al. 1992). 
Their Hbeta radial trend is also steeper than reported here.
An important caveat to these comparisons is that the luminosity range spanned by the 
Carter et al. sample is larger than that used here, or by Guzm\'an et al. We have already commented
that the radial effects could be larger for galaxies of lower velocity dispersion, which would work in the right direction to reconcile
the two studies. 

Pimbblet et al. (2006) recently measured a radial shift in the $B-R$ colour of the red sequence, in 11 clusters at 
$z\sim0.1$. The fitted sample spans a luminosity range comparable to NFPS. The magnitude of the colour trend is 
$-0.011\pm0.003$\,mag per $h^{-1}_{50}$\,Mpc. Given also their measurement of a significant rising radial trend in the H$\delta$ 
absorption, Pimbblet et al. argue this 
is due to younger populations at larger radius. Since their definition of H$\delta$ is not the same as the HdF index
employed here, we compare on the basis of the $B-R$ colour gradient predicted from our age, metallicity and \afe\ trends. 
Using models by Maraston (1998, 2005), our radial age gradient would yield $\delta(B-R) = -0.029\pm0.008$ blueing at the virial 
radius, while the metallicity effect contributes  $\delta(B-R) = +0.003\pm0.006$ reddening. The Maraston models do not account for \afe\
variations; using a differential estimate of $\frac{\partial(B-R)}{\partial[\alpha/\rm Fe]}\approx0.12$ derived from 
Salasnich et al. (2000), the reduced \afe\ at large radius further reddens the outer galaxies by $+0.006\pm0.001$. The net
effect is thus  $\delta(B-R) = -0.020\pm0.010$ at the virial radius. The median $R_{200}$ for the Pimbblet et al. sample is 
2.9\,$h^{-1}_{50}$\,Mpc. Thus our prediction for their colour trend is $-0.007\pm0.003$\,mag per $h^{-1}_{50}$\,Mpc, which 
is consistent with their result. 

In summary, our results from NFPS strongly confirm previous hints of radial differences in 
stellar populations, while our more general approach helps to resolve apparently contradictory
claims concerning the driving parameters of the trends.

\subsection{Comparison with field vs cluster studies}

Arguably a more natural way to search for the influence of high-density 
environments is to compare galaxies in rich clusters to those 
in the `field'. The caveat to this, of course, is that many alternative definitions
of `field' are in use, e.g. large distance to nearest neighbour; found in a large-scale
under-density; not in a known rich cluster. As an example, an elliptical member of a compact group 
could qualify under some of these criteria, but not under others. 

It is tempting to compare the radial trends of galaxy populations in clusters with the 
results from field surveys, since galaxies falling into the cluster for the first time 
will not yet have experienced any effect of the cluster, and should match the field population. 
However, this comparison should be treated with some caution, since 
many galaxies may have been accreted into their host cluster as members of infalling groups, 
rather than as true field objects. If so, the outer cluster members may already have been substantially 
`pre-processed' in the group environment (e.g. Kodama et al. 2001), and
should not be expected to match the field samples. 

Many works have employed spectroscopic indices to test for systematic differences in the 
stellar populations of early-type galaxies in the field and in clusters. 
For example, Bernardi et al. (1998) found a small offset towards lower Mg$_2$
in the field, by 0.007$\pm$0.002\,mag at given velocity dispersion, 
and interpreted this as an age difference with the field galaxies 1\,Gyr
younger than cluster populations.  Kuntschner et al. (2002) used multiple indices, reporting
younger ages, by 2--3\,Gyr compared to clusters, for a small sample of very isolated galaxies.
Thomas et al. (2005) obtained a similar offset of $\sim$2\,Gyr between field and cluster ellipticals. 
None of the above studies found evidence for an environmental shift in \afe, although both Kuntschner
et al. and Thomas et al. report an increased total metallicity \zh, which is not reproduced in our 
results. 

Most recently, Bernardi et al. (2006) analysed composite spectra derived from the Sloan Digital Sky Survey, for 
early-type galaxies selected by morphology and spectral type. 
They conclude that early-type galaxies in the lowest-density environments are younger by $\sim$1\,Gyr, 
than their counterparts of similar mass in the densest regions. The study finds a shift of 0.02\,dex in \afe, 
with relatively less $\alpha$-enhancement in the field, but no significant offset is found for the overall metallicity. 
These results are qualitatively in striking agreement with the conclusions of this paper, 
if we allow that our outer cluster galaxies should approximately resemble their low-density population.

\subsection{Comparison with galaxy formation models}

The observed pattern of cluster-centric gradients agrees qualitatively with a model in which galaxies 
in the outer parts of cluster had more extended star formation activity, leading both to younger ages and more
pollution of the interstellar gas by iron from SNe Ia, before star formation ceased. 
Thomas et al. (2005) constructed simple models of chemical enrichment in an extended burst,  
where the star-formation rate is Gaussian with FWHM $\tau$ in Gyr, to describe the relationship between
\afe\ and the formation time-scale $\tau$. Adopting their relation, \afe$\approx\frac{1}{5}-\frac{1}{6}\log\tau$, 
our shift of $\Delta$\afe$=0.05$ between the core and the virial radius corresponds to 
$\Delta\log\tau=6\times\Delta[\alpha/{\rm Fe}]\approx0.3$. Thus, galaxies at $R_{200}$ formed their stars on 
time-scales twice as long as those for the inner galaxies. 

In high-resolution dark-matter simulations of cluster formation, sub-haloes found
near the cluster centre at the present day were accreted at larger redshift. For instance Gao et al. (2004) 
find average infall redshifts of $z\approx0.4$ at the virial radius, compared to $z\approx1$ in the core. 
Together with the expectation that star-formation in the sub-haloes may be truncated after 
accretion to the cluster, perhaps through removal or exhaustion of the gas reservoir (Larson et al. 1980; Balogh et al. 2000), 
this correlation will lead to an age gradient in the observed sense. Given a longer period of continuous star-formation 
prior to accretion, outer galaxies will also tend to have less $\alpha$-enriched stellar populations. 

As a toy model, we consider at face value the average infall redshifts of Gao et al.'s Figure 15 (lower-left panel),
and convolve with a star-formation rate which is constant until accretion, then declines exponentially with
an e-folding time of 1 Gyr. The model yields present day luminosity-weighted ages of 10\,Gyr in the centre, 
and 7\,Gyr at the (physical) virial radius. Clearly a number of important factors have been 
neglected here:  First, the average infall redshift shown by Gao et al. gives the time since the sub-haloes entered the
{\it ultimate} parent halo. In practice, some of these will have been absorbed within smaller parent haloes 
at earlier epochs, so that truncation may have taken place at a higher redshift. Second, as emphasized by Kodama \& Bower (2001),
the evolution of the global star-formation rate implies that the typical field accreted galaxy 
at $z\approx0.4$ was less active than that accreted at $z\approx1$ (Madau, Pozzetti \& Dickinson 1998). Finally, and more trivially, 
the predicted gradients are in terms of physical radius from the cluster centre, rather than projected radius as observed. 
Each of these three effects leads to an over-estimation of the trends, so that the 30\% age change from core to virial 
radius should be regarded as an upper limit on the likely gradient. 

Cosmological N-body simulations and semi-analytic galaxy formation models can provide
more realistic descriptions of both the star-formation history and the merger history of the accreted galaxies. 
With the large mass resolution afforded by the Virgo Consortium's Millennium Simulation (Springel et al. 2005), 
it is possible to follow sub-haloes corresponding to sub-$L^\star$ cluster members, and analyse their properties
with reference to their location in the parent cluster. Recently, De Lucia et al. (2006) applied
updated `Munich' semi-analytic recipes to Millennium Simulation merger trees, to model
the formation of elliptical galaxies. They predict that the luminosity-weighted
stellar age declines from $\sim$12\,Gyr at the cluster centres to $\sim$10.5\,Gyr at
$R_{200}$, a $\sim$13\% effect (their Figure 8a). In comparing to our observational results, 
two caveats should be borne in mind. First, we measure a trend in projected radius, 
whereas De Lucia et al. plot the physical distance. Second, De Lucia et al. 
do not correct for the overall age-versus-mass trend, and comment that some of their cluster-centric 
gradient arises from mass segregation. Examining their Figures 8a and 8c, most of the
change in age is from 0.1\,$R_{200}$ to 1.0\,$R_{200}$, a range over which the median stellar
mass of ellipticals changes by only $\sim$2\%. Even considering the inner 0.1\,$R_{200}$, 
the 10\% mass change, combined with their weak age--mass trend for massive galaxies (their Figure 6a),
cannot be primarily responsible for their radial gradient. Thus, despite the difficulties in matching 
the observed results to the models, our measured age shift of $15\pm4\%$ appears to be compatible
with these predictions. In future work, we intend to make a more detailed confrontation with the 
newly-updated `Durham' semi-analytic predictions as described by Bower et al. (2005). 

In summary, a modest age--radius (and \afe--radius) relation is expected in hierarchical models, 
given the ample observational evidence for a quenching of star formation in the infall regions 
of nearby clusters (Lewis et al. 2002; G\'omez et al. 2003). Our detection of a radial age trend 
in the red sequence, of the expected sense and order of magnitude, suggests that present-day red-sequence 
galaxies are the descendents of similar quenching mechanisms occurring at earlier cosmic epochs. 

\section{Conclusions}\label{sec:concs}

We have presented an analysis of the absorption line-strength indices of red-sequence galaxies in clusters,
focussing on the subtle effects of environment, as indicated by distance from the host cluster centre. 

Our principal conclusions can be summarised as follows: 

\begin{enumerate}

\item
The line-strengths of red-sequence galaxies exhibit systematic 
trends as a function of distance from the centre of their host cluster, after
controlling for the dominant trend with velocity dispersion. 

\item
The cluster-centric trends are small compared to the overall changes
of line-strengths with velocity dispersion, and small compared to the total
scatter around these relations. None the less, for some indices the
radial effects are highly significant. 

\item
The pattern of gradients is different from that of the index versus velocity 
dispersion relations. The radial trends are not a scaled version of the mass trends. 
Whatever the underlying process driving the radial differences, 
it is distinct from that which imposes the overall mass sequence. 

\item
There is a segregation in the indices between those which respond strongly to age or \afe\ ratio,
where the radial trends are opposite in sign to the mass trends, and the \zh-indicators, which
have radial trends of the same sign as the mass trends. 

\item
Within the context of the TMBK stellar population models, the cluster-centric 
gradients are best described by a significant variation in age and relative $\alpha$-abundance.
The sense is that red-sequence galaxies at one virial radius, $R_{200}$, on average have younger stars by $15\pm4$\% than
red-sequence galaxies of the same mass at the cluster centre. The $\alpha$ abundance, relative to iron, is
$10\pm2$\% lower at $R_{200}$ than at the cluster centre, again at fixed mass. 
The overall metallicity \zh\ does not show a significant trend.

\item
Available morphology data suggests that although stellar populations appear to depend strongly on
bulge-fraction $B/T$, this does not translate into a significant contribution to the radial trend. 
The radial behaviour is not driven by a changing morphological mix, but rather a change in typical
stellar populations for galaxies of the same morphology, as well as the same mass. 

\item
Robustness tests using subsets of the input data suggest that the radial trend signals are quite evenly 
divided between distant and nearby clusters, and between the two celestial hemispheres. Using only the relaxed
clusters with well-defined centres, we recover slightly stronger trends. 
There is a suggestion that the radial trend in age is strongest for galaxies of lower mass.

\item
The age trend determined here from the observed line-strength indices is in the sense generically
predicted from hierarchical clustering with star-formation truncation in clusters. Numerically, 
there is good agreement with very recent predictions from semi-analytic galaxy formation models
(De Lucia et al. 2006), which predict an age decline of $\sim$13\% out to $R_{200}$. 

\end{enumerate}

In summary, we have shown that the average star formation histories of red-sequence galaxies depend on
their location in the cluster, in addition to the overall dependence on galaxy mass. Our results lend 
support to the notion that the red sequence was established through the accretion of star-forming galaxies 
from the field, with a subsequent environmental truncation of star formation. Together
with our earlier measurement of the mass dependence of the average star-formation history (Nelan et al. 2005), 
a picture is beginning to emerge for how the red sequence became populated as a cosmic history progressed. 
Further progress in this area will be driven by improved spectroscopic observations of faint 
cluster members, which appear to exhibit stronger signatures of later accretion. Additional insights should 
result from comparing in greater detail with morphological and photometric evidence for galaxy transformation in clusters
(e.g. Stevenson et al. 2006), with observations at higher redshifts (e.g. Tanaka et al. 2005), 
and with improved modelling of the hierarchical galaxy formation process.

\section*{Acknowledgments}

RJS was supported for part of this work from the PPARC rolling grant `Extragalactic Astronomy \& Cosmology at Durham 2005--2010' (C.S. Frenk). 
MJH acknowledges support from the NSERC of Canada. JEN was supported by a Dartmouth College Fellowship, a Graduate Assistance in Areas of 
National Need fellowship, and a NASA space grant. We gratefully acknowledge the substantial assignment of NOAO observing resources to the NFPS program. 
This publication makes use of data products from the Two Micron All Sky Survey, which is a joint project of the University of Massachusetts and the 
Infrared Processing and Analysis Center/California Institute of Technology, funded by the National Aeronautics and Space Administration and the 
National Science Foundation.

{}

\label{lastpage}
\end{document}